\documentclass{article}
\usepackage{eurosym}
\usepackage{algorithm}
\usepackage{amsfonts}
\usepackage{setspace}
\usepackage{wrapfig}
\usepackage{subfig}
\usepackage{graphicx}
\usepackage[margin=1in]{geometry}
\usepackage{float}
\usepackage[toc,page]{appendix}
\usepackage{color}
\usepackage{amsmath}
\usepackage{setspace}
\usepackage{float}
\usepackage{amssymb}

\newtheorem{theorem}{Theorem}
\newtheorem{acknowledgement}[theorem]{Acknowledgement}

\DeclareMathOperator{\SR}{SR}
\DeclareMathOperator{\DUR}{DUR}
\input{tcilatex}
\begin{document}

\title{A closed-form solution for optimal mean-reverting trading strategies}
\author{ Alexander Lipton\thanks{%
The Jerusalem School of Business Administration, The Hebrew University of
Jerusalem, Jerusalem, Israel; Connection Science and Engineering,
Massachusetts Institute of Technology, Cambridge, MA, USA; Investimizer,
Chicago, IL, USA; SilaMoney, Portland, OR, USA. E-mail: alexlipt@mit.edu} \
and Marcos Lopez de Prado\thanks{%
Operations Research and Information Engineering, Cornell University, New
York, NY, USA; Investimizer, Chicago, IL, USA; True Positive Technologies,
New York, NY, USA. E-mail ml863@cornell.edu }}
\date{}
\maketitle

\begin{abstract}
When prices reflect all available information, they oscillate around an
equilibrium level. This oscillation is the result of the temporary market
impact caused by waves of buyers and sellers. This price behavior can be
approximated through an Ornstein-Uhlenbeck (OU) process.

Market makers provide liquidity in an attempt to monetize this oscillation.
They enter a long position when a security is priced below its estimated
equilibrium level, and they enter a short position when a security is priced
above its estimated equilibrium level. They hold that position until one of
three outcomes occur: (1) they achieve the targeted profit; (2) they
experience a maximum tolerated loss; (3) the position is held beyond a
maximum tolerated horizon.

All market makers are confronted with the problem of defining profit-taking
and stop-out levels. More generally, all execution traders acting on behalf
of a client must determine at what levels an order must be fulfilled. Those
optimal levels can be determined by maximizing the trader's Sharpe ratio in
the context of OU processes via Monte Carlo experiments, \cite{Lopez2018}.
This paper develops an analytical framework and derives those optimal levels
by using the method of heat potentials, \cite{Lipton2018, Lipton2020a}.
\end{abstract}

\section{Introduction}

Mean-reverting trading strategies in various contexts have been studied for
decades, see, e.g., \cite{Vidyamurthy2004, Gatev2006, Govender2011,
Gregory2011, Krauss2015, Leung2015a, Leung2015b, Li2015, Huck2015,
Goncu2016, Suzuki2018}. For instance, Elliott \textit{et al.} explained how
mean-reverting processes might be used in pairs trading and developed
several methods for parameter estimation, \cite{Elliott2005}. Avellaneda and
Lee used mean-reverting processes for pairs trading, and modeled the hitting
time to find the exit rule of the trade, \cite{Avellaneda2010}. Bertram
developed some analytic formulae for statistical arbitrage trading where the
security price follows an Ornstein--Uhlenbeck (O-U) process, \cite%
{Bertram2009, Bertram2010}. Lindberg and his coauthors model the spread
between two assets as an O-U process and study the optimal liquidation
strategy for an investor who wants to optimize profit over the opportunity
cost, \cite{Ekstrom2010, Ekstrom2011, Larsson2013, Lindberg2014}. Lopez de
Prado (Chapter 13) considered trading rules for discrete-time mean-reverting
trading strategies and found optimal trading rules using Monte Carlo
simulations, \cite{Lopez2018}.

By its very nature, the energy market is particularly well suited to
mean-reverting trading strategies. Numerous researchers discuss these
strategies, see, e.g., \cite{Cummins2012, Baviera2017, Liu2017}, among
others.

Usually, it is assumed that the stochastic process underlying mean-reverting
trading strategies is the standard O-U process. However, in practice, jumps
do play a major role. Hence, some attention had been devoted to L\'{e}%
vy-driven O-U processes, see, e.g., \cite{Larsson2013, Goncu2016, Endres2017}%
. Although Endres and St\"{u}binger, \cite{Endres2017}, present a fairly
detailed exposition, their central equation (9) is incorrect because it
ignores the possibility of a L\'{e}vy-driven O-U process to overshoot the
chosen boundaries.

We emphasize that most, if not all, analytical results derived by the above
authors, are asymptotic and valid for perpetual trading strategies only,
see, e.g., \cite{Bertram2009, Bertram2010, Ekstrom2010, Ekstrom2011,
Lataillade2012, Larsson2013, Lindberg2014, Zeng2014, Bai2018}. While
interesting from a theoretical standpoint, they have limited application in
practice. In contrast, our approach deals with finite maturity trading
strategies, and, because of that, has immediate applications.

When prices reflect all available information, they oscillate around an
equilibrium level. This oscillation is the result of the temporary market
impact caused by waves of buyers and sellers. The resulting price behavior
can be approximated through an O-U process. The parameters of the process
might be estimated using historical data.

Market makers provide liquidity in an attempt to monetize this oscillation.
They enter a long position when a security is priced below its estimated
equilibrium level, and they enter a short position when a security is priced
above its estimated equilibrium level. They hold that position until one of
three outcomes occurs: (A) they achieve a targeted profit; (B) they
experience a maximum tolerated loss; (C) the position is held beyond a
maximum tolerated horizon.

All traders are confronted with the problem of defining profit-taking and
stop-out levels. More generally, all execution traders acting on behalf of a
client must determine at what levels an order must be fulfilled. Lopez de
Prado (Chapter 13) explains how to identify those optimal levels in the
sense of maximizing the trader's Sharpe ratio ($\SR$) in the context of O-U
processes via Monte Carlo experiments, \cite{Lopez2018}. Although Lopez de
Prado (p. 192) conjectured the existence of an analytical solution to this
problem, he identified it as an open problem. In this paper, we solve the
critical question of finding optimal trading rules analytically by using the
method of heat potentials. These optimal profit-taking/stop-loss trading
rules for mean-reverting trading strategies provide the algorithm that must
be followed to exit a position. To put it differently, we find the optimal
exit corridor to maximize the $\SR$ of the strategy.

The method of heat potential is a highly powerful and versatile approach
popular in mathematical physics; see, e.g., \cite{Tikhonov1963,
Rubinstein1971, Kartashov2001, Wilson2012} among others. It has been
successfully used in numerous important fields, such as thermal engineering,
nuclear engineering, and material science. However, it is not particularly
popular in mathematical finance,even though the first important use case was
given by Lipton almost twenty years ago. Specifically, Lipton considered
pricing barrier options with curvilinear barriers, see \cite{Lipton2001},
Section 12.2.3, pp. 462--467. More recently, Lipton and Kaushansky described
several important financial applications of the method, see \cite%
{Lipton2018, Lipton2019, Lipton2020a, Lipton2020b}.

The $\SR$ is defined as the ratio between the expected returns of an
execution algorithm and the standard deviation of the same returns. The
returns are computed as the logarithmic ratio between the exit and entry
prices, times the sign of the order side ($-1$ for a sell order, $+1$ for a
buy order). Our choice of the $\SR$ as an objective function is due to two
reasons: (A) The $\SR$ is the most popular criterion for investment
efficiency, \cite{Lopez2013}; (B) The $\SR$ can be understood as a t-value
of the estimated gains, and modelled accordingly for inferential purposes.
The distributional properties of the $\SR$ are well-known, and this
statistic can be deflated when the assumption of normality is violated, \cite%
{Lopez2014}.

Having an analytical estimation of the optimal profit-taking and stop-out
levels allows traders to deploy tactical execution algorithms, with maximal
expected $\SR$. Rather than deriving an \textquotedblleft
all-weather\textquotedblright\ execution algorithm, which supposedly works
under every market regime, traders can use our analytical solution for
deploying the algorithm that maximizes the $\SR$ under the prevailing market
regime, \cite{Lopez2019}.

\section{Definitions of variables}

Suppose an investment strategy $S$ invests in $i=1,...I$ opportunities or
bets. At each opportunity $i$, $S$ takes a position of $m_{i}$ units of
security $X$, where $m_{i}\in (-\infty ,\infty )$. The transaction that
entered such opportunity was priced at a value $m_{i}P_{i,0}$, where $%
P_{i,0} $ is the average price per unit at which the $m_{i}$ securities were
transacted. As other market participants transact security $X$, we can
mark-to-market (MtM) the value of that opportunity $i$ after $t$ observed
transactions as $m_{i}P_{i,t}$. This represents the value of opportunity $i$
if it were liquidated at the price observed in the market after $t$
transactions. Accordingly, we can compute the MtM profit/loss of opportunity 
$i$ after $t$ transactions as $\pi _{i,t}=m_{i}(P_{i,t}-P_{i,0})$.

A standard trading rule provides the logic for exiting opportunity $i$ at $%
t=T_{i}$. This occurs as soon as one of two conditions is verified:

\begin{itemize}
\item $\pi _{i,T_{i}}\geq \overline{\pi }$, where $\overline{\pi }>0$ is the
profit-taking threshold.

\item $\pi _{i,T_{i}}\leq \underline{\pi }$, where $\underline{\pi }<0$ is
the \ stop-loss threshold.
\end{itemize}

Because $\underline{\pi }<\overline{\pi }$, one and only one of the two exit
conditions can trigger the exit from opportunity $i$. Assuming that
opportunity $i$ can be exited at $T_{i}$, its final profit/loss is $\pi
_{i,T_{i}}$. At the \ onset \ of each opportunity, the \ goal \ is to
realize an expected profit 
\begin{equation*}
\begin{array}{c}
\mathbb{E}_{0}[\pi _{i,T_{i}}]=m_{i}(\mathbb{E}_{0}[P_{i,T_{i}}]-P_{i,0}),%
\end{array}%
\end{equation*}%
where \ $\mathbb{E}_{0}[P_{i,T_{i}}]$ is the forecasted price and $P_{i,0}$
is the entry level of opportunity $i$.

\section{Parameter estimation}

Consider the discrete O-U process on a price series $\left\{ P_{i,t}\right\} 
$:%
\begin{equation*}
\begin{array}{c}
P_{i,t}-\mathbb{E}_{0}[P_{i,T_{i}}]=\kappa \left(
E_{0}[P_{i,T_{i}}]-P_{i,t-1}\right) +\sigma \varepsilon _{i,t},%
\end{array}%
\end{equation*}%
such that the random shocks are IID distributed $\varepsilon _{i,t}\sim 
\mathcal{N}\left( 0,1\right) $. The seed value for this process is $P_{i,0}$%
, the level targeted by opportunity $i$ is $\mathbb{E}_{0}[P_{i,T_{i}}]$,
and $\kappa $ determines the speed at which $P_{i,0}\ $converges towards $%
\mathbb{E}_{0}[P_{i,T_{i}}]$.

We estimate the input parameters $\left\{ \kappa ,\sigma \right\} $, by
stacking the opportunities as:%
\begin{equation*}
\begin{array}{c}
X=\left( \mathbb{E}_{0}[P_{0,T_{0}}]-P_{0,0},\mathbb{E}%
_{0}[P_{0,T_{0}}]-P_{0,1},...,\mathbb{E}_{0}[P_{0,T_{0}}]-P_{0,T-1},...,%
\mathbb{E}_{0}[P_{I,T_{I}}]-P_{I,0},...,\mathbb{E}%
_{0}[P_{I,T_{I}}]-P_{I,T-1}\right) ^{T}, \\ 
\\ 
Y=\left( P_{0,1}-\mathbb{E}_{0}[P_{0,T_{0}}],P_{0,2}-\mathbb{E}%
_{0}[P_{0,T_{0}}],...,P_{0,T}-\mathbb{E}_{0}[P_{0,T_{0}}],...,P_{I,1}-%
\mathbb{E}_{0}[P_{I,T_{I}}],...,P_{I,T}-\mathbb{E}_{0}[P_{I,T_{I}}]\right)
^{T},%
\end{array}%
\end{equation*}%
where $\left( ...\right) ^{T}$ denotes vector transposition. Applying OLS on
the above equation, we can estimate the original O-U parameters as follows:%
\begin{equation*}
\begin{array}{c}
\hat{\kappa}=\frac{\mathrm{cov}\left[ Y,X\right] }{\mathrm{cov}\left[ X,X%
\right] },\ \ \ \hat{\xi}=Y-\hat{\kappa}X,\ \ \ \hat{\sigma}=\sqrt{\mathrm{%
cov}\left[ \hat{\xi},\hat{\xi}\right] },%
\end{array}%
\end{equation*}%
where, as usual, $\mathrm{cov}\left[ .,.\right] $ is the covariance
operator. We use the above estimations to find optimal stop-loss and
take-profit bounds.

\section{Explicit problem formulation}

In this rather technical section, we perform transformations in order to
formulate the problem in terms of heat potentials.

Consider a long investment strategy $S$ and suppose profit/loss opportunity
is driven by an O-U\ process (see \cite{Lopez2018} among many others):%
\begin{equation}
\begin{array}{c}
dx^{\prime }=\kappa ^{\prime }\left( \theta ^{\prime }-x^{\prime }\right)
dt^{\prime }+\sigma ^{\prime }dW_{t^{\prime }},\ \ \ x^{\prime }\left(
0\right) =0,%
\end{array}
\label{OUProcess}
\end{equation}%
and a trading rule $R=\{\underline{\pi }^{\prime },\overline{\pi }^{\prime
},T^{\prime }\}$, $\underline{\pi }^{\prime }<0$, $\overline{\pi }^{\prime
}>0$. It is important to understand what are the natural units associated
with the O-U process (\ref{OUProcess}). To this end we can use its
steady-state. The steady-state expectation of the above process is $\theta $%
, while its standard deviation is given by $\Omega ^{\prime }=\sigma
^{\prime }/\sqrt{2\kappa ^{\prime }}$.

As usual, an appropriate scaling is helpful to remove superfluous
parameters. To this end, we define%
\begin{equation*}
\begin{array}{c}
t=\kappa ^{\prime }t^{\prime },\ \ \ \ T=\kappa ^{\prime }T^{\prime },\ \ \
\ x=\frac{\sqrt{\kappa ^{\prime }}}{\sigma ^{\prime }}x^{\prime },\ \ 
\overline{\pi }=\frac{\sqrt{\kappa ^{\prime }}}{\sigma ^{\prime }}\overline{%
\pi }^{\prime },\ \ \ \underline{\pi }=\frac{\sqrt{\kappa ^{\prime }}}{%
\sigma ^{\prime }}\underline{\pi }^{\prime },\ \ \ \theta =\frac{\sqrt{%
\kappa ^{\prime }}}{\sigma ^{\prime }}\theta ^{\prime }\ ,\ \ \ E=\frac{%
E^{\prime }}{\sqrt{\kappa ^{\prime }}\sigma ^{\prime }},\ \ \ F=\frac{%
F^{\prime }}{\kappa ^{\prime }\sigma ^{\prime 2}},%
\end{array}%
\end{equation*}%
and get%
\begin{equation*}
\begin{array}{c}
dx=\left( \theta -x\right) dt+dW_{t},%
\end{array}%
\end{equation*}%
in the domain%
\begin{equation*}
\ 
\begin{array}{c}
\underline{\pi }\leq x\leq \overline{\pi },\ \ \ 0\leq t\leq T.%
\end{array}%
\end{equation*}%
The steady-state distribution has the expectation of $\theta $, and the
standard deviation $\Omega =1/\sqrt{2}$.

According to the trading rule, we exit the trade either when: (A) the price
hits $\overline{\pi }$ to take profit; (B) the price hits $\underline{\pi }$
to stop losses; (C) the trade expires at $t=T$. For a short investment
strategy, the roles of $\{\underline{\pi },\overline{\pi }\}$ are reversed -
profits equal to $-\underline{\pi }$ are taken when when the price hits $%
\underline{\pi }$, and losses equal $-\overline{\pi }$ are realized when the
price hits $\overline{\pi }$. Given the fact that the reflection $%
x\rightarrow -x$ leaves the initial condition unchanged and transforms the
original O-U process into the O-U\ process of the form%
\begin{equation*}
\begin{array}{c}
dx=\left( -\theta -x\right) dt+dW_{t},\ \ \ x\left( 0\right) =0,%
\end{array}%
\end{equation*}%
we can restrict ourselves to the case $\theta \geq 0$. More explicitly and
intuitively, we go long when $\theta \geq 0$ and short when $\theta <0$.
Assuming that we know the trading rule $\{\underline{\pi }\left( \theta
,T\right) ,\overline{\pi }\left( \theta ,T\right) ,T\}$ for $\theta \geq 0$,
the corresponding trading rule for $\theta <0$ has the form%
\begin{equation*}
\begin{array}{c}
\{\underline{\pi }\left( \theta ,T\right) ,\overline{\pi }\left( \theta
,T\right) ,T\}=\{-\underline{\pi }\left( -\theta ,T\right) ,-\overline{\pi }%
\left( -\theta ,T\right) ,T\}.%
\end{array}%
\end{equation*}%
Thus, we are interested in the maximization of the $\SR$ for nonnegative $%
\theta \geq 0$. We formulate this mathematically below.

For a given $T$, we define the stopping time $\iota =\inf \{t:x_{t}=%
\overline{\pi }\text{ or }x_{t}=\underline{\pi }$ or $t=T\}$. We wish to
determine optimal $\overline{\pi }>0,\underline{\pi }<0$,to maximize the $%
\SR
$,%
\begin{equation*}
\begin{array}{c}
\SR=\frac{\mathbb{E}\left\{ x_{\iota }/\iota \right\} }{\sqrt{\mathbb{E}%
\left\{ x_{\iota }^{2}/\iota ^{2}\right\} -\left( \mathbb{E}\left\{ x_{\iota
}/\iota \right\} \right) ^{2}}},%
\end{array}%
\end{equation*}%
We also need to know the expected duration of the trade,%
\begin{equation*}
\begin{array}{c}
\DUR=\mathbb{E}\left\{ \iota \right\} .%
\end{array}%
\end{equation*}

In order to calculate the corresponding $\SR$ and $\DUR$ we proceed as
follows. We solve three terminal boundary value problems (TBVPs) of the form%
\begin{equation*}
\begin{array}{c}
E_{t}\left( t,x\right) +\left( \theta -x\right) E_{x}\left( t,x\right) +%
\frac{1}{2}E_{xx}\left( t,x\right) =0, \\ 
\\ 
E\left( t,\overline{\pi }\right) =\frac{\overline{\pi }}{t},\ \ \ E\left( t,%
\underline{\pi }\right) =\frac{\underline{\pi }}{t}, \\ 
\\ 
E\left( T,x\right) =\frac{x}{T},%
\end{array}%
\end{equation*}%
\begin{equation*}
\begin{array}{c}
F_{t}\left( t,x\right) +\left( \theta -x\right) F_{x}\left( t,x\right) +%
\frac{1}{2}F_{xx}\left( t,x\right) =0, \\ 
\\ 
F\left( t,\overline{\pi }\right) =\frac{\overline{\pi }^{2}}{t^{2}},\ \ \
F\left( t,\underline{\pi }\right) =\frac{\underline{\pi }^{2}}{t^{2}}, \\ 
\\ 
F\left( T,x\right) =\frac{x^{2}}{T^{2}},%
\end{array}%
\end{equation*}%
and%
\begin{equation*}
\begin{array}{c}
G_{t}\left( t,x\right) +\left( \theta -x\right) G_{x}\left( t,x\right) +%
\frac{1}{2}G_{xx}\left( t,x\right) =0, \\ 
\\ 
G\left( t,\overline{\pi }\right) =t,\ \ \ G\left( t,\underline{\pi }\right)
=t, \\ 
\\ 
G\left( T,x\right) =T,%
\end{array}%
\end{equation*}%
We represent the $\SR$ and $\DUR$ as%
\begin{equation*}
\begin{array}{c}
\SR=\frac{E\left( 0,0\right) }{\sqrt{F\left( 0,0\right) -\left( E\left(
0,0\right) \right) ^{2}}},%
\end{array}%
\end{equation*}%
\begin{equation*}
\begin{array}{c}
\DUR=G\left( 0,0\right) .%
\end{array}%
\end{equation*}

We wish to use the method of heat potentials to solve the above TBVPs.
First, we define 
\begin{equation*}
\begin{array}{c}
\tau =T-t,%
\end{array}%
\end{equation*}%
and get initial boundary value problems (IBVPs):%
\begin{equation*}
\begin{array}{c}
E_{\tau }\left( \tau ,x\right) =\left( \theta -x\right) E_{x}\left( \tau
,x\right) +\frac{1}{2}E_{xx}\left( \tau ,x\right) , \\ 
\\ 
E\left( \tau ,\overline{\pi }\right) =\frac{\overline{\pi }}{\left( T-\tau
\right) },\ \ \ E\left( \tau ,\underline{\pi }\right) =\frac{\underline{\pi }%
}{\left( T-\tau \right) }, \\ 
\\ 
E\left( 0,x\right) =\frac{x}{T},%
\end{array}%
\end{equation*}%
\begin{equation*}
\begin{array}{c}
F_{\tau }\left( \tau ,x\right) =\left( \theta -x\right) F_{x}\left( \tau
,x\right) +\frac{1}{2}F_{xx}\left( \tau ,x\right) , \\ 
\\ 
F\left( \tau ,\overline{\pi }\right) =\frac{\overline{\pi }^{2}}{\left(
T-\tau \right) ^{2}},\ \ \ F\left( \tau ,\underline{\pi }\right) =\frac{%
\underline{\pi }^{2}}{\left( T-\tau \right) ^{2}}, \\ 
\\ 
F\left( 0,x\right) =\frac{x^{2}}{T^{2}},%
\end{array}%
\end{equation*}%
\begin{equation*}
\begin{array}{c}
G_{\tau }\left( \tau ,x\right) =\left( \theta -x\right) G_{x}\left( \tau
,x\right) +\frac{1}{2}G_{xx}\left( \tau ,x\right) , \\ 
\\ 
G\left( \tau ,\overline{\pi }\right) =\left( T-\tau \right) ,\ \ \ G\left(
\tau ,\underline{\pi }\right) =\left( T-\tau \right) , \\ 
\\ 
G\left( 0,x\right) =T,%
\end{array}%
\end{equation*}%
\begin{equation*}
\begin{array}{c}
\SR=\frac{E\left( T,0\right) }{\sqrt{F\left( T,0\right) -\left( E\left(
T,0\right) \right) ^{2}}},%
\end{array}%
\end{equation*}%
\begin{equation*}
\begin{array}{c}
\DUR=G\left( T,0\right) .%
\end{array}%
\end{equation*}

Second, we define 
\begin{equation*}
\begin{array}{c}
\upsilon =\frac{1-e^{-2\tau }}{2},\ \ \ \xi =e^{-\tau }\left( x-\theta
\right) ,%
\end{array}%
\end{equation*}%
so that%
\begin{equation*}
\begin{array}{c}
\partial _{\tau }=\left( 1-2\upsilon \right) \partial _{\upsilon }-\xi
\partial _{\xi },\ \ \ \partial _{x}=\sqrt{1-2\upsilon }\partial _{\xi }.%
\end{array}%
\end{equation*}%
Accordingly,%
\begin{equation*}
\begin{array}{c}
E_{\upsilon }\left( \upsilon ,\xi \right) =\frac{1}{2}E_{\xi \xi }\left(
\upsilon ,\xi \right) , \\ 
\\ 
E\left( \upsilon ,\overline{\Pi \left( \upsilon \right) }\right) =\frac{2%
\overline{\pi }}{\ln \left( \frac{\left( 1-2\upsilon \right) }{\left(
1-2\Upsilon \right) }\right) },\ \ \ \ \ E\left( \upsilon ,\underline{\Pi
\left( \upsilon \right) }\right) =\frac{2\underline{\pi }}{\ln \left( \frac{%
\left( 1-2\upsilon \right) }{\left( 1-2\Upsilon \right) }\right) }, \\ 
\\ 
E\left( 0,\xi \right) =-\frac{2\left( \xi +\theta \right) }{\ln \left(
1-2\Upsilon \right) },%
\end{array}%
\end{equation*}%
\begin{equation*}
\begin{array}{c}
F_{\upsilon }\left( \upsilon ,\xi \right) =\frac{1}{2}F_{\xi \xi }\left(
\upsilon ,\xi \right) , \\ 
\\ 
F\left( \upsilon ,\overline{\Pi \left( \upsilon \right) }\right) =\frac{4%
\overline{\pi }^{2}}{\left( \ln \left( \frac{\left( 1-2\upsilon \right) }{%
\left( 1-2\Upsilon \right) }\right) \right) ^{2}},\ \ \ \ \ F\left( \upsilon
,\underline{\Pi \left( \upsilon \right) }\right) =\frac{\underline{\pi }^{2}%
}{\left( \ln \left( \frac{\left( 1-2\upsilon \right) }{\left( 1-2\Upsilon
\right) }\right) \right) ^{2}}, \\ 
\\ 
\ F\left( 0,\xi \right) =\frac{4\left( \xi +\theta \right) ^{2}}{\left( \ln
\left( 1-2\Upsilon \right) \right) ^{2}},%
\end{array}%
\end{equation*}%
\begin{equation*}
\begin{array}{c}
G_{\upsilon }\left( \upsilon ,\xi \right) =\frac{1}{2}G_{\xi \xi }\left(
\upsilon ,\xi \right) , \\ 
\\ 
G\left( \upsilon ,\overline{\Pi \left( \upsilon \right) }\right) =\frac{1}{2}%
\ln \left( \frac{\left( 1-2\upsilon \right) }{\left( 1-2\Upsilon \right) }%
\right) ,\ \ \ \ \ G\left( \upsilon ,\underline{\Pi \left( \upsilon \right) }%
\right) =\frac{1}{2}\ln \left( \frac{\left( 1-2\upsilon \right) }{\left(
1-2\Upsilon \right) }\right) , \\ 
\\ 
G\left( 0,\xi \right) =-\frac{1}{2}\ln \left( 1-2\Upsilon \right) ,%
\end{array}%
\end{equation*}%
\begin{equation*}
\begin{array}{c}
\SR=\frac{E\left( \Upsilon ,\varpi \right) }{\sqrt{F\left( \Upsilon ,\varpi
\right) -\left( E\left( \Upsilon ,\varpi \right) \right) ^{2}}},%
\end{array}%
\end{equation*}%
\begin{equation*}
\begin{array}{c}
\DUR=G\left( \Upsilon ,\varpi \right) .%
\end{array}%
\end{equation*}%
Here 
\begin{equation*}
\begin{array}{c}
\Upsilon =\frac{1-e^{-2T}}{2},~\ \ \ \varpi =-\sqrt{1-2\Upsilon }\theta , \\ 
\\ 
\overline{\Pi }(\upsilon )=\sqrt{1-2\upsilon }\left( \overline{\pi }-\theta
\right) ,\ \ \ \underline{\Pi }(\upsilon )=\sqrt{1-2\upsilon }\left( 
\underline{\pi }-\theta \right) .%
\end{array}%
\end{equation*}%
As usual, we have to account for the initial conditions. To this end, we
introduce%
\begin{equation*}
\begin{array}{c}
\hat{E}\left( \upsilon ,\xi \right) =E\left( \upsilon ,\xi \right) +\frac{%
2\left( \xi +\theta \right) }{\ln \left( 1-2\Upsilon \right) }, \\ 
\\ 
\hat{F}\left( \upsilon ,\xi \right) =F\left( \upsilon ,\xi \right) -\frac{%
4\left( \upsilon +\left( \xi +\theta \right) ^{2}\right) }{\left( \ln \left(
1-2\Upsilon \right) \right) ^{2}}, \\ 
\\ 
\hat{G}\left( \upsilon ,\xi \right) =G\left( \upsilon ,\xi \right) +\frac{1}{%
2}\ln \left( 1-2\Upsilon \right) ,%
\end{array}%
\end{equation*}%
where%
\begin{equation*}
\begin{array}{c}
\hat{E}_{\upsilon }\left( \upsilon ,\xi \right) =\frac{1}{2}\hat{E}_{\xi \xi
}\left( \upsilon ,\xi \right) , \\ 
\\ 
\hat{E}\left( \upsilon ,\underline{\Pi }\left( \upsilon \right) \right) =%
\underline{e}\left( \upsilon \right) ,\ \ \ \hat{E}\left( \upsilon ,%
\overline{\Pi }(\upsilon )\right) =\overline{e}\left( \upsilon \right) , \\ 
\\ 
\hat{E}\left( 0,\xi \right) =0,%
\end{array}%
\end{equation*}%
\begin{equation*}
\begin{array}{c}
\hat{F}_{\upsilon }\left( \upsilon ,\xi \right) =\frac{1}{2}\hat{F}_{\xi \xi
}\left( \upsilon ,\xi \right) , \\ 
\\ 
\hat{F}\left( \upsilon ,\underline{\Pi }\left( \upsilon \right) \right) =%
\underline{f}\left( \upsilon \right) ,\ \ \ \hat{F}\left( \upsilon ,%
\overline{\Pi }(\upsilon )\right) =\overline{f}\left( \upsilon \right) , \\ 
\\ 
\hat{F}\left( 0,\xi \right) =0,%
\end{array}%
\end{equation*}%
\begin{equation*}
\begin{array}{c}
\hat{G}_{\upsilon }\left( \upsilon ,\xi \right) =\frac{1}{2}\hat{G}_{\xi \xi
}\left( \upsilon ,\xi \right) , \\ 
\\ 
\ \ \hat{G}\left( \upsilon ,\underline{\Pi }\left( \upsilon \right) \right) =%
\underline{g}\left( \upsilon \right) ,\ \ \ \hat{G}\left( \upsilon ,%
\overline{\Pi }(\upsilon )\right) =\overline{g}\left( \upsilon \right) , \\ 
\\ 
\hat{G}\left( 0,\xi \right) =0,%
\end{array}%
\end{equation*}%
where%
\begin{equation*}
\begin{array}{c}
\underline{e}\left( \upsilon \right) =\frac{2\underline{\pi }}{\ln \left( 
\frac{\left( 1-2\upsilon \right) }{\left( 1-2\Upsilon \right) }\right) }+%
\frac{2(\underline{\Pi }(\upsilon )+\theta )}{\ln \left( 1-2\Upsilon \right) 
},\ \ \ \overline{e}\left( \upsilon \right) =\frac{2\overline{\pi }}{\ln
\left( \frac{\left( 1-2\upsilon \right) }{\left( 1-2\Upsilon \right) }%
\right) }+\frac{2(\overline{\Pi }(\upsilon )+\theta )}{\ln \left(
1-2\Upsilon \right) } \\ 
\\ 
\underline{f}\left( \upsilon \right) =\frac{4\underline{\pi }^{2}}{\left(
\ln \left( \frac{\left( 1-2\upsilon \right) }{\left( 1-2\Upsilon \right) }%
\right) \right) ^{2}}-\frac{4\left( \upsilon +(\underline{\Pi }\left(
\upsilon \right) +\theta )^{2}\right) }{\left( \ln \left( 1-2\Upsilon
\right) \right) ^{2}},\ \ \ \ \overline{f}\left( \upsilon \right) =\frac{4%
\overline{\pi }^{2}}{\left( \ln \left( \frac{\left( 1-2\upsilon \right) }{%
\left( 1-2\Upsilon \right) }\right) \right) ^{2}}-\frac{4\left( \upsilon +(%
\overline{\Pi }(\upsilon )+\theta )^{2}\right) }{\left( \ln \left(
1-2\Upsilon \right) \right) ^{2}}, \\ 
\\ 
\underline{g}\left( \upsilon \right) =\frac{1}{2}\ln \left( 1-2\upsilon
\right) ,\ \ \ \overline{g}\left( \upsilon \right) =\frac{1}{2}\ln \left(
1-2\upsilon \right) .%
\end{array}%
\end{equation*}%
Accordingly,%
\begin{equation}
\begin{array}{c}
\SR=\frac{\hat{E}\left( \Upsilon ,\varpi \right) -\frac{2\left( \varpi
+\theta \right) }{\ln \left( 1-2\Upsilon \right) }}{\sqrt{\hat{F}\left(
\Upsilon ,\varpi \right) -\left( \hat{E}\left( \Upsilon ,\varpi \right)
\right) ^{2}+\frac{4\left( \Upsilon +\ln \left( 1-2\Upsilon \right) \left(
\varpi +\theta \right) \hat{E}\left( \Upsilon ,\varpi \right) \right) }{%
\left( \ln \left( 1-2\Upsilon \right) \right) ^{2}}}},%
\end{array}
\label{sharpe_final_eq}
\end{equation}%
\begin{equation*}
\begin{array}{c}
\DUR=\hat{G}\left( \Upsilon ,\varpi \right) -\frac{1}{2}\ln \left(
1-2\Upsilon \right) .%
\end{array}%
\end{equation*}

After the above transformations are performed, the problem becomes solvable
by the method of heat potentials.

\section{The method of heat potentials}

Now we are ready to use the classical method of heat potentials to calculate
the $\SR$. Consider $\hat{E}$. We have to solve the following coupled system
of Volterra integral equations:%
\begin{equation}
\begin{array}{c}
\underline{\varepsilon }\left( \upsilon \right) +\frac{1}{\sqrt{2\pi }}%
\dint\limits_{0}^{\upsilon }\frac{\left( \underline{\Pi }\left( \upsilon
\right) -\underline{\Pi }\left( \zeta \right) \right) e^{-\frac{\left( 
\underline{\Pi }\left( \upsilon \right) -\underline{\Pi }\left( \zeta
\right) \right) ^{2}}{2\left( \upsilon -\zeta \right) }}}{\left( \upsilon
-\zeta \right) ^{3/2}}\underline{\varepsilon }\left( \zeta \right) d\zeta \\ 
\\ 
+\frac{1}{\sqrt{2\pi }}\dint\limits_{0}^{\upsilon }\frac{\left( \underline{%
\Pi }\left( \upsilon \right) -\overline{\Pi }\left( \zeta \right) \right)
e^{-\frac{\left( \underline{\Pi }\left( \upsilon \right) -\overline{\Pi }%
\left( \zeta \right) \right) ^{2}}{2\left( \upsilon -\zeta \right) }}}{%
\left( \upsilon -\zeta \right) ^{3/2}}\overline{\varepsilon }\left( \zeta
\right) d\zeta =\underline{e}\left( \upsilon \right) ,%
\end{array}
\label{pdf1}
\end{equation}%
\begin{equation}
\begin{array}{c}
-\overline{\varepsilon }\left( \upsilon \right) +\frac{1}{\sqrt{2\pi }}%
\dint\limits_{0}^{\upsilon }\frac{\left( \overline{\Pi }\left( \upsilon
\right) -\underline{\Pi }\left( \zeta \right) \right) e^{-\frac{\left( 
\overline{\Pi }\left( \upsilon \right) -\underline{\Pi }\left( \zeta \right)
\right) ^{2}}{2\left( \upsilon -\zeta \right) }}}{\left( \upsilon -\zeta
\right) ^{3/2}}\underline{\varepsilon }\left( \zeta \right) d\zeta \\ 
\\ 
+\frac{1}{\sqrt{2\pi }}\dint\limits_{0}^{\upsilon }\frac{\left( \overline{%
\Pi }\left( \upsilon \right) -\overline{\Pi }\left( \zeta \right) \right)
e^{-\frac{\left( \overline{\Pi }\left( \upsilon \right) -\overline{\Pi }%
\left( \zeta \right) \right) ^{2}}{2\left( \upsilon -\zeta \right) }}}{%
\left( \upsilon -\zeta \right) ^{3/2}}\overline{\varepsilon }\left( \zeta
\right) d\zeta =\overline{e}\left( \upsilon \right) ,%
\end{array}
\label{pdf2}
\end{equation}%
Once these equations are solved, $\hat{E}\left( \upsilon ,\xi \right) $ can
be written as follows:%
\begin{equation}
\begin{array}{c}
\hat{E}\left( \upsilon ,\xi \right) =\frac{1}{\sqrt{2\pi }}%
\dint\limits_{0}^{\upsilon }\frac{\left( \xi -\underline{\Pi }\left( \zeta
\right) \right) e^{-\frac{\left( \xi -\underline{\Pi }\left( \zeta \right)
\right) ^{2}}{2\left( \upsilon -\zeta \right) }}}{\left( \upsilon -\zeta
\right) ^{3/2}}\underline{\varepsilon }\left( \zeta \right) d\zeta +\frac{1}{%
\sqrt{2\pi }}\dint\limits_{0}^{\upsilon }\frac{\left( \xi -\overline{\Pi }%
\left( \zeta \right) \right) e^{-\frac{\left( \xi -\overline{\Pi }\left(
\zeta \right) \right) ^{2}}{2\left( \upsilon -\zeta \right) }}}{\left(
\upsilon -\zeta \right) ^{3/2}}\overline{\varepsilon }\left( \zeta \right)
d\zeta .%
\end{array}
\label{EqE}
\end{equation}

We can find $\hat{F}\left( \upsilon ,\xi \right) $ by the same token:%
\begin{equation}
\begin{array}{c}
\underline{\phi }\left( \upsilon \right) +\frac{1}{\sqrt{2\pi }}%
\dint\limits_{0}^{\upsilon }\frac{\left( \underline{\Pi }\left( \upsilon
\right) -\underline{\Pi }\left( \zeta \right) \right) e^{-\frac{\left( 
\underline{\Pi }\left( \upsilon \right) -\underline{\Pi }\left( \zeta
\right) \right) ^{2}}{2\left( \upsilon -\zeta \right) }}}{\left( \upsilon
-\zeta \right) ^{3/2}}\underline{\phi }\left( \zeta \right) d\zeta \\ 
\\ 
+\frac{1}{\sqrt{2\pi }}\dint\limits_{0}^{\upsilon }\frac{\left( \underline{%
\Pi }\left( \upsilon \right) -\overline{\Pi }\left( \zeta \right) \right)
e^{-\frac{\left( \underline{\Pi }\left( \upsilon \right) -\overline{\Pi }%
\left( \zeta \right) \right) ^{2}}{2\left( \upsilon -\zeta \right) }}}{%
\left( \upsilon -\zeta \right) ^{3/2}}\overline{\phi }\left( \zeta \right)
d\zeta =\underline{f}\left( \upsilon \right) ,%
\end{array}
\label{phi1}
\end{equation}%
\begin{equation}
\begin{array}{c}
-\overline{\phi }\left( \upsilon \right) +\frac{1}{\sqrt{2\pi }}%
\dint\limits_{0}^{\upsilon }\frac{\left( \overline{\Pi }\left( \upsilon
\right) -\underline{\Pi }\left( \zeta \right) \right) e^{-\frac{\left( 
\overline{\Pi }\left( \upsilon \right) -\underline{\Pi }\left( \zeta \right)
\right) ^{2}}{2\left( \upsilon -\zeta \right) }}}{\left( \upsilon -\zeta
\right) ^{3/2}}\underline{\phi }\left( \zeta \right) d\zeta \\ 
\\ 
+\frac{1}{\sqrt{2\pi }}\dint\limits_{0}^{\upsilon }\frac{\left( \overline{%
\Pi }\left( \upsilon \right) -\overline{\Pi }\left( \zeta \right) \right)
e^{-\frac{\left( \overline{\Pi }\left( \upsilon \right) -\overline{\Pi }%
\left( \zeta \right) \right) ^{2}}{2\left( \upsilon -\zeta \right) }}}{%
\left( \upsilon -\zeta \right) ^{3/2}}\overline{\phi }\left( \zeta \right)
d\zeta =\overline{f}\left( \upsilon \right) ,%
\end{array}
\label{phi2}
\end{equation}%
\begin{equation}
\begin{array}{c}
\hat{F}\left( \upsilon ,\xi \right) =\frac{1}{\sqrt{2\pi }}%
\dint\limits_{0}^{\upsilon }\frac{\left( \xi -\underline{\Pi }\left( \zeta
\right) \right) e^{-\frac{\left( \xi -\underline{\Pi }\left( \zeta \right)
\right) ^{2}}{2\left( \upsilon -\zeta \right) }}}{\left( \upsilon -\zeta
\right) ^{3/2}}\underline{\phi }\left( \zeta \right) d\zeta +\frac{1}{\sqrt{%
2\pi }}\dint\limits_{0}^{\upsilon }\frac{\left( \xi -\overline{\Pi }\left(
\zeta \right) \right) e^{-\frac{\left( \xi -\overline{\Pi }\left( \zeta
\right) \right) ^{2}}{2\left( \upsilon -\zeta \right) }}}{\left( \upsilon
-\zeta \right) ^{3/2}}\overline{\phi }\left( \zeta \right) d\zeta .%
\end{array}
\label{EqF}
\end{equation}

In particular,%
\begin{equation*}
\begin{array}{c}
\hat{E}\left( \Upsilon ,\varpi \right) =\frac{1}{\sqrt{2\pi }}%
\dint\limits_{0}^{\Upsilon }\frac{\left( \varpi -\underline{\Pi }\left(
\zeta \right) \right) e^{-\frac{\left( \varpi -\underline{\Pi }\left( \zeta
\right) \right) ^{2}}{2\left( \Upsilon -\zeta \right) }}}{\left( \Upsilon
-\zeta \right) ^{3/2}}\underline{\varepsilon }\left( \zeta \right) d\zeta +%
\frac{1}{\sqrt{2\pi }}\dint\limits_{0}^{\Upsilon }\frac{\left( \varpi -%
\overline{\Pi }\left( \zeta \right) \right) e^{-\frac{\left( \varpi -%
\overline{\Pi }\left( \zeta \right) \right) ^{2}}{2\left( \Upsilon -\zeta
\right) }}}{\left( \Upsilon -\zeta \right) ^{3/2}}\overline{\varepsilon }%
\left( \zeta \right) d\zeta , \\ 
\\ 
\hat{F}\left( \Upsilon ,\varpi \right) =\frac{1}{\sqrt{2\pi }}%
\dint\limits_{0}^{\Upsilon }\frac{\left( \varpi -\underline{\Pi }\left(
\zeta \right) \right) e^{-\frac{\left( \varpi -\underline{\Pi }\left( \zeta
\right) \right) ^{2}}{2\left( \Upsilon -\zeta \right) }}}{\left( \Upsilon
-\zeta \right) ^{3/2}}\underline{\phi }\left( \zeta \right) d\zeta +\frac{1}{%
\sqrt{2\pi }}\dint\limits_{0}^{\Upsilon }\frac{\left( \varpi -\overline{\Pi }%
\left( \zeta \right) \right) e^{-\frac{\left( \varpi -\overline{\Pi }\left(
\zeta \right) \right) ^{2}}{2\left( \Upsilon -\zeta \right) }}}{\left(
\Upsilon -\zeta \right) ^{3/2}}\overline{\phi }\left( \zeta \right) d\zeta .%
\end{array}%
\end{equation*}%
It is important to notice that $\left( \underline{\varepsilon }\left( \zeta
\right) ,\overline{\varepsilon }\left( \zeta \right) \right) $ and $\left( 
\underline{\phi }\left( \zeta \right) ,\overline{\phi }\left( \zeta \right)
\right) $ are singular at $\zeta =\Upsilon $. However, due to the dampening
impact of the exponents $\exp \left( -\left( \varpi -\overline{\Pi }\left(
\zeta \right) \right) ^{2}/2\left( \Upsilon -\zeta \right) \right) $, the
corresponding integrals still converge.

We now know $\hat{E}\left( \Upsilon ,\varpi \right) ,\hat{F}\left( \Upsilon
,\varpi \right) $ and calculate the $\SR$ by using Eq. (\ref{sharpe_final_eq}%
). $\hat{G}\left( \Upsilon ,\varpi \right) $ and $\DUR$ can be calculated in
a similar fashion.

\section{Numerical method}

\label{sec:numerical} To compute the $\SR$, we need to find $\hat{E}%
(\Upsilon ,\varpi )$ and $\hat{F}(\Upsilon ,\varpi )$, and then apply Eq. (%
\ref{sharpe_final_eq}). $\hat{E}(\Upsilon ,\varpi )$ and $\hat{F}(\Upsilon
,\varpi )$ can be computed using Eqs (\ref{EqE}) and (\ref{EqF}) by simple
integration with pre-computed $\left( \underline{\varepsilon },\overline{%
\varepsilon }\right) $ and $\left( \underline{\phi },\overline{\phi }\right) 
$. In this section, we develop a numerical method to compute these
quantities by solving Eqs (\ref{pdf1})--(\ref{pdf2}), and (\ref{phi1})--(\ref%
{phi2}) by extending the methods described in Lipton and Kaushansky, \cite%
{Lipton2018, Lipton2020a}. For illustrative purposes we develop a simple
scheme based on the trapezoidal rule for Stieltjes integrals.

We want to solve a generic system of the form: 
\begin{equation*}
\begin{array}{c}
\nu ^{1}(\upsilon )+\int_{0}^{\upsilon }\frac{K^{1,1}(\upsilon ,s)}{\sqrt{%
\upsilon -s}}\nu ^{1}(s)\,ds+\int_{0}^{\upsilon }K^{1,2}(\upsilon ,s)\nu
^{2}(s)\,ds=\chi ^{1}(\upsilon ), \\ 
\\ 
-\nu ^{2}(\upsilon )+\int_{0}^{\upsilon }K^{2,1}(\upsilon ,s)\nu
^{1}(s)\,ds+\int_{0}^{\upsilon }\frac{K^{2,2}(\upsilon ,s)}{\sqrt{\upsilon -s%
}}\nu ^{2}(s)\,ds=\chi ^{2}(\upsilon ),%
\end{array}%
\end{equation*}%
with respect to variables $(\nu ^{1}(\upsilon ),\nu ^{2}(\upsilon ))$, where 
\begin{equation*}
\begin{array}{c}
K^{1,1}(\upsilon ,s)=\frac{1}{\sqrt{2\pi }}\frac{\underline{\Pi }(\upsilon )-%
\underline{\Pi }(s)}{\upsilon -s}\exp \left( -\frac{(\underline{\Pi }%
(\upsilon )-\underline{\Pi }(s))^{2}}{2(\upsilon -s)}\right) , \\ 
\\ 
K^{1,2}(\upsilon ,s)=\frac{1}{\sqrt{2\pi }}\frac{\underline{\Pi }(\upsilon )-%
\overline{\Pi }(s)}{\left( \upsilon -s\right) ^{3/2}}\exp \left( -\frac{(%
\underline{\Pi }(\upsilon )-\overline{\Pi }(s))^{2}}{2(\upsilon -s)}\right) ,
\\ 
\\ 
K^{2,1}(\upsilon ,s)=\frac{1}{\sqrt{2\pi }}\frac{\overline{\Pi }(\upsilon )-%
\underline{\Pi }(s)}{\left( \upsilon -s\right) ^{3/2}}\exp \left( -\frac{(%
\overline{\Pi }(\upsilon )-\underline{\Pi }(s))^{2}}{2(\upsilon -s)}\right) ,
\\ 
\\ 
K^{2,2}(\upsilon ,s)=\frac{1}{\sqrt{2\pi }}\frac{\overline{\Pi }(\upsilon )-%
\overline{\Pi }(s)}{\upsilon -s}\exp \left( -\frac{(\overline{\Pi }(\upsilon
)-\overline{\Pi }(s))^{2}}{2(\upsilon -s)}\right) .%
\end{array}%
\end{equation*}%
It is clear that 
\begin{equation*}
\begin{array}{c}
K^{1,1}\left( \upsilon ,\upsilon \right) =\frac{1}{\sqrt{2\pi }}%
\lim_{s\rightarrow \upsilon }\frac{\underline{\Pi }(\upsilon )-\underline{%
\Pi }(s)}{\upsilon -s}=\frac{\theta -\underline{\pi }}{\sqrt{2\pi }\sqrt{%
1-2\upsilon }}, \\ 
\\ 
K^{1,2}\left( \upsilon ,\upsilon \right) =0, \\ 
\\ 
K^{2,1}\left( \upsilon ,\upsilon \right) =0, \\ 
\\ 
K^{2,2}\left( \upsilon ,\upsilon \right) =\frac{1}{\sqrt{2\pi }}%
\lim_{s\rightarrow \upsilon }\frac{\overline{\Pi }(\upsilon )-\overline{\Pi }%
(s)}{\upsilon -s}=\frac{\theta -\overline{\pi }}{\sqrt{2\pi }\sqrt{%
1-2\upsilon }}.%
\end{array}%
\end{equation*}

We can equally rewrite the relevant integrals as Stieltjes integrals 
\begin{equation*}
\begin{array}{c}
\nu ^{1}(\upsilon )-2\int_{0}^{\upsilon }K^{1,1}(\upsilon ,s)\nu ^{1}(s)\,d%
\sqrt{\upsilon -s}+\int_{0}^{\upsilon }K^{1,2}(\upsilon ,s)\nu
^{2}(s)\,ds=\chi ^{1}(\upsilon ), \\ 
\\ 
-\nu ^{2}(\upsilon )+\int_{0}^{\upsilon }K^{2,1}(\upsilon ,s)\nu
^{1}(s)\,ds-2\int_{0}^{\upsilon }K^{2,2}(\upsilon ,s)\nu ^{2}(s)\,d\sqrt{%
\upsilon -s}=\chi ^{2}(\upsilon ).%
\end{array}%
\end{equation*}%
Consider a grid $0=\upsilon _{0}<\upsilon _{1}<\ldots <\upsilon
_{n}=\Upsilon $, and let $\Delta _{k,l}=\upsilon _{k}-\upsilon _{l}$. Then,
using the trapezoidal rule for approximation of integrals, we get the
following approximation of last two equations: 
\begin{equation*}
\begin{array}{c}
\nu _{k}^{1}+\sum_{i=1}^{k}\left( \frac{\left( K_{k,i}^{1,1}\nu
_{i}^{1}+K_{k,i-1}^{1,1}\nu _{i-1}^{1}\right) }{\left( \sqrt{\Delta _{k,i}}+%
\sqrt{\Delta _{k,i-1}}\right) }+\frac{1}{2}\left( K_{k,i}^{1,2}\nu
_{i}^{2}+K_{k,i-1}^{1,2}\nu _{i-1}^{2}\right) \right) \Delta _{i,i-1}=\chi
_{k}^{1}, \\ 
\\ 
-\nu _{k}^{2}+\sum_{i=1}^{k}\left( \frac{1}{2}\left( K_{k,i}^{2,1}\nu
_{i}^{1}+K_{k,i-1}^{2,1}\nu _{i-1}^{1}\right) +\frac{\left( K_{k,i}^{2,2}\nu
_{i}^{2}+K_{k,i-1}^{2,2}\nu _{i-1}^{2}\right) }{\left( \sqrt{\Delta _{k,i}}+%
\sqrt{\Delta _{k,i-1}}\right) }\right) \Delta _{i,i-1}=\chi _{k}^{2}.%
\end{array}%
\end{equation*}%
where%
\begin{equation*}
\begin{array}{c}
\nu _{i}^{\alpha }=\nu ^{\alpha }\left( \upsilon _{i}\right) ,\ \ \ \ \ \chi
_{i}^{\alpha }=\chi ^{\alpha }\left( \upsilon _{i}\right) ,\ \ \ \ \
K_{k,j}^{\alpha ,\beta }=K^{\alpha ,\beta }(\upsilon _{k},\upsilon _{i}),\ \
\ \ \ \alpha ,\beta =1,2.%
\end{array}%
\end{equation*}%
Taking into account that 
\begin{equation*}
\begin{array}{c}
\left( \nu _{0}^{1},\nu _{0}^{2}\right) =\left( \chi _{0}^{1},-\chi
_{0}^{2}\right) ,\ \ \ \left( \nu _{1}^{1},\nu _{1}^{2}\right) =\left( \frac{%
\chi _{1}^{1}}{\left( 1+K_{1,1}^{1,1}\sqrt{\upsilon _{1}}\right) },-\frac{%
\chi _{1}^{2}}{\left( 1-K_{1,1}^{2,2}\sqrt{\upsilon _{1}}\right) }\right) ,%
\end{array}%
\end{equation*}%
and assuming that $\left( \nu _{2}^{1},\nu _{2}^{2}\right) ,\ldots ,\left(
\nu _{k-1}^{1},\nu _{k-1}^{2}\right) $ have been computed, we can easily
find $\left( \nu _{k}^{1},\nu _{k}^{2}\right) $:%
\begin{equation*}
\begin{array}{c}
\nu _{k}^{1}=\left( 1+K_{k,k}^{1,1}\sqrt{\Delta _{k,k-1}}\right) ^{-1}\left(
\chi _{k}^{1}-K_{k,k-1}^{1,1}\nu _{k-1}^{1}\sqrt{\Delta _{k,k-1}}-\frac{1}{2}%
K_{k,k-1}^{1,2}\nu _{k-1}^{2}\Delta _{k,k-1}\right. \\ 
\\ 
\left. -\sum_{i=1}^{k-1}\left( \frac{\left( K_{k,i}^{1,1}\nu
_{i}^{1}+K_{k,i-1}^{1,1}\nu _{i-1}^{1}\right) }{\left( \sqrt{\Delta _{k,i}}+%
\sqrt{\Delta _{k,i-1}}\right) }+\frac{1}{2}\left( K_{k,i}^{1,2}\nu
_{i}^{2}+K_{k,i-1}^{1,2}\nu _{i-1}^{2}\right) \right) \Delta _{i,i-1}\right)
, \\ 
\nu _{k}^{2}=\left( -1+K_{k,k}^{2,2}\sqrt{\Delta _{k,k-1}}\right)
^{-1}\left( \chi _{k}^{2}-\frac{1}{2}K_{k,k-1}^{2,1}\nu _{k-1}^{1}\Delta
_{k,k-1}-K_{k,k-1}^{2,2}\nu _{k-1}^{2}\sqrt{\Delta _{k,k-1}}\right. \\ 
\\ 
\left. -\sum_{i=1}^{k-1}\left( \frac{1}{2}\left( K_{k,i}^{2,1}\nu
_{i}^{1}+K_{k,i-1}^{2,1}\nu _{i-1}^{1}\right) +\frac{\left( K_{k,i}^{2,2}\nu
_{i}^{2}+K_{k,i-1}^{2,2}\nu _{i-1}^{2}\right) }{\left( \sqrt{\Delta _{k,i}}+%
\sqrt{\Delta _{k,i-1}}\right) }\right) \Delta _{i,i-1}\right) .%
\end{array}%
\end{equation*}

The approximation error of the integrals is of order $O(\Delta ^{2})$, where 
$\Delta =\max_{i}\Delta _{i,i-1})$. Hence, on uniform grid, the convergence
is of order $O(\Delta )$. We emphasize that, due to the nature of $\left( 
\underline{e}(\upsilon ),\overline{e}(\upsilon )\right) $, etc., it is
necessary to use a highly inhomogeneous grid which is concentrated near the
right endpoint.

\subsection{Computation of the Sharpe ratio}

Once $\left( \underline{\varepsilon }(\upsilon ),\overline{\varepsilon }%
(\upsilon )\right) $ and $\left( \underline{\phi }(\upsilon ),\overline{\phi 
}(\upsilon )\right) $ are computed, we can approximate $\hat{E}(\upsilon
,\xi )$ and $\hat{F}(\upsilon ,\xi )$. We interested to compute these
functions at one point $(\Upsilon ,\varpi )$, which can be done by
approximation of the integrals using the trapezoidal rule: 
\begin{equation}
\begin{array}{c}
\hat{E}(\Upsilon ,\varpi )=\frac{1}{2}\sum_{i=1}^{k}\left( \underline{w}%
_{n,i}\underline{\varepsilon }_{i}+\underline{w}_{n,i-1}\underline{%
\varepsilon }_{i-1}+\ \overline{w}_{n,i}\overline{\varepsilon }_{i}+\ 
\overline{w}_{n,i-1}\overline{\varepsilon }_{i-1}\right) \Delta _{i,i-1},%
\end{array}
\label{Eeqapprox}
\end{equation}%
and%
\begin{equation}
\begin{array}{c}
\hat{F}(\Upsilon ,\varpi )=\frac{1}{2}\sum_{i=1}^{k}\left( \underline{w}%
_{n,i}\underline{\phi }_{i}+\underline{w}_{n,i-1}\underline{\phi }_{i-1}+\ 
\overline{w}_{n,i}\overline{\phi }_{i}+\ \overline{w}_{n,i-1}\overline{\phi }%
_{i-1}\right) \Delta _{i,i-1}.%
\end{array}
\label{Feqapprox}
\end{equation}%
The corresponding weights are as follows:%
\begin{equation*}
\begin{array}{ccc}
\underline{w}_{n,i}=\frac{\left( \varpi -\underline{\Pi }_{i}\right) e^{-%
\frac{\left( \varpi -\underline{\Pi }_{i}\right) ^{2}}{2\Delta _{n,i}}}}{%
\sqrt{2\pi }\Delta _{n,i}^{3/2}}, & \overline{w}_{n,i}=\frac{\left( \varpi -%
\overline{\Pi }_{i}\right) e^{-\frac{\left( \varpi -\overline{\Pi }%
_{i}\right) ^{2}}{2\Delta _{n,i}}}}{\sqrt{2\pi }\Delta _{n,i}^{3/2}} & 1\leq
i<n, \\ 
\underline{w}_{n,i}=0 & \overline{w}_{n,i}=0, & i=n.%
\end{array}%
\end{equation*}%
As a result, we get the following algorithm for the numerical evaluation of
the $\SR$.

\begin{algorithm}
\caption{Numerical evaluation of the Sharpe ratio}
\begin{tabular}{cc}
Step 1 & Define a time grid $0=\upsilon_{0}<\upsilon_{1}<\ldots <\Upsilon$. \\
Step 2 & Compute $\underline{\epsilon}(\upsilon), \overline{\epsilon}(\upsilon), \underline{\phi}(\upsilon), \overline{\phi}(\upsilon)$ using numerical method in Section \ref{sec:numerical}. \\
Step 3 & Compute $\hat{E}(\Upsilon,\varpi)$ by using (\ref{Eeqapprox}). \\
Step 4 & Compute $\hat{F}(\Upsilon,\varpi )$ by using Eq. (\ref{Feqapprox}).\\
Step 5 & Compute the Sharpe ratio by using Eq. (\ref{sharpe_final_eq}). \\
\end{tabular}
\label{Algo1}
\end{algorithm}

\section{Numerical results}

\subsection{Comparison with Monte Carlo simulations}

We compute the $\SR$ for various values of $\underline{\pi }$ and $\overline{%
\pi }$, and as a result show the $\SR$ as a function of $(\underline{\pi },%
\overline{\pi })$. After that one can choose $(\underline{\pi },\overline{%
\pi })$ in order to maximize the $\SR$.

\begin{figure}[tbp]
\begin{center}
\subfloat[]{\includegraphics[width=0.5\textwidth]
{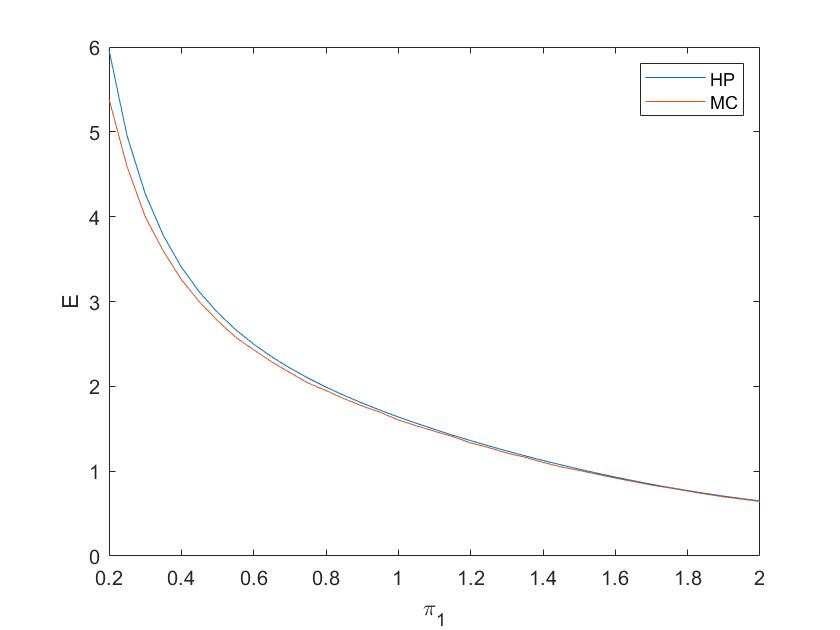}}
\subfloat[]{\includegraphics[width=0.5\textwidth]
{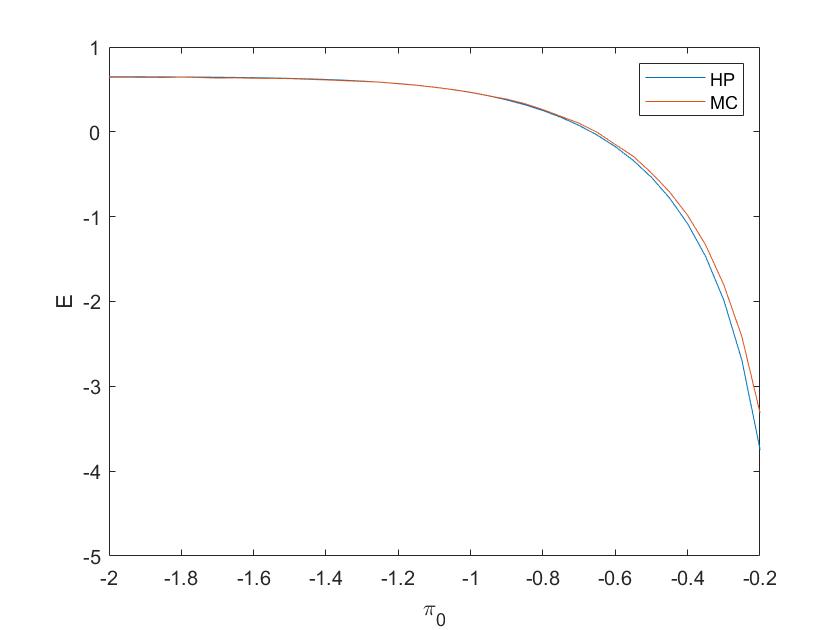}}%
\\[0pt]
\subfloat[]{\includegraphics[width=0.5\textwidth]
{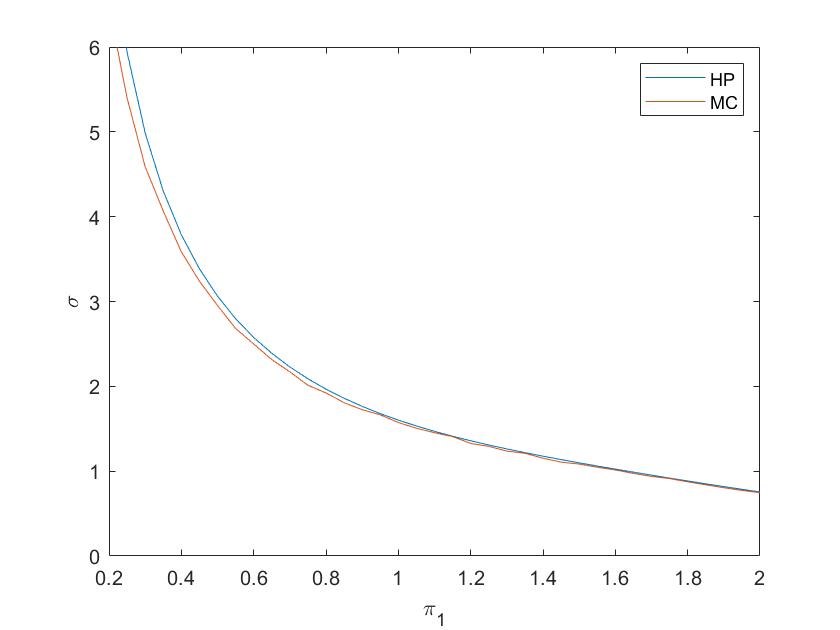}}
\subfloat[]{\includegraphics[width=0.5\textwidth]
{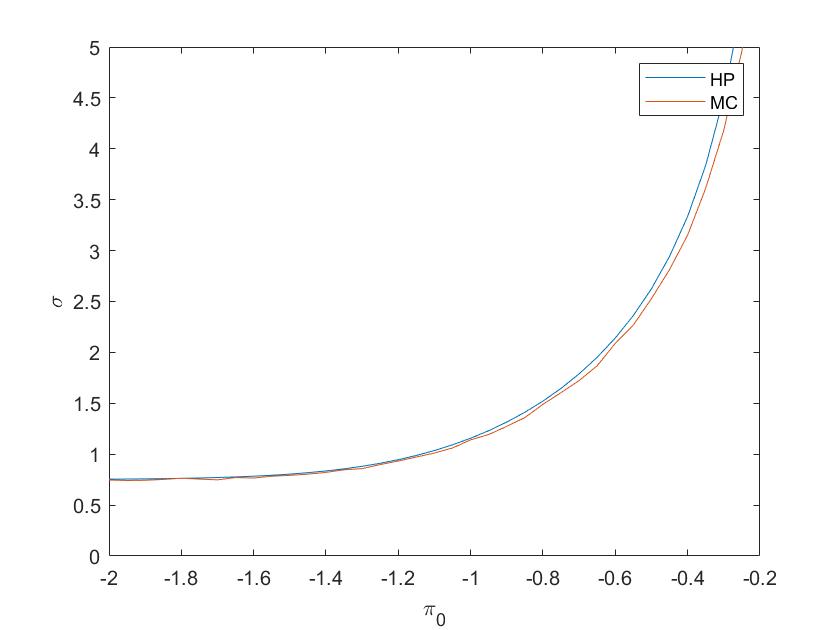}}%
\\[0pt]
\subfloat[]{\includegraphics[width=0.5\textwidth]
{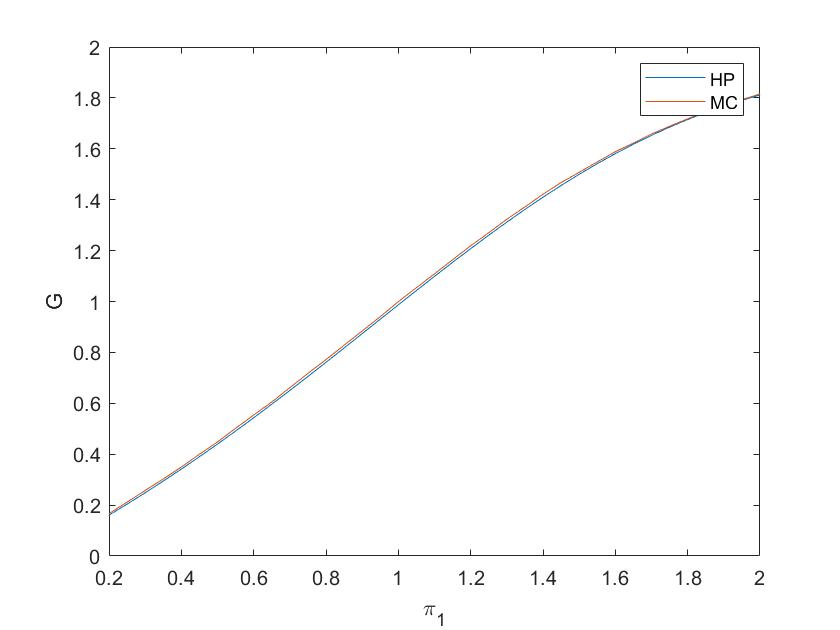}}
\subfloat[]{\includegraphics[width=0.5\textwidth]
{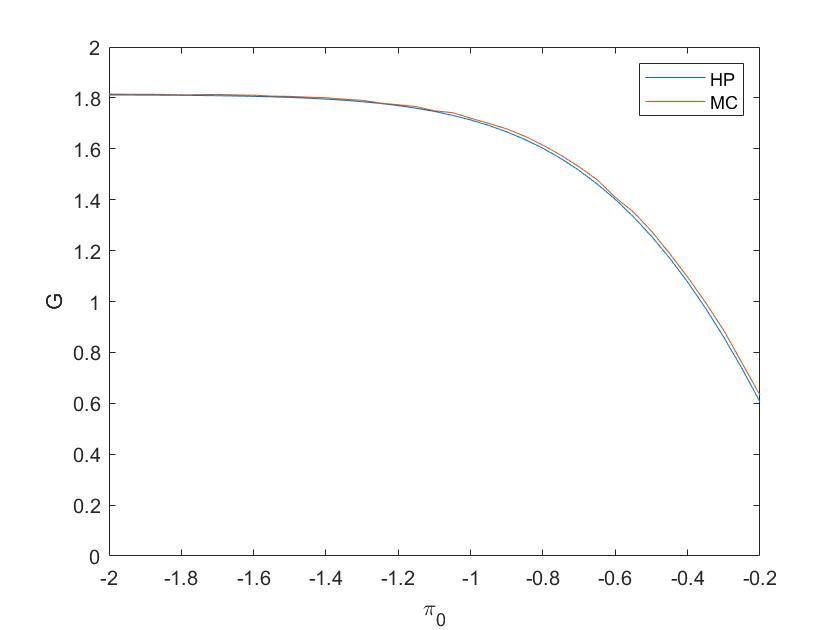}}
\end{center}
\par
\vspace{-10pt}
\caption{ (a) $E,\protect\sigma =\protect\sqrt{F-E^{2}}$ and $G$ as
functions of $\overline{\protect\pi }\equiv \protect\pi _{1}$ computed by
using the method of heat potentials and the Monte Carlo method for $\protect%
\underline{\protect\pi }=-2$; (b) Same quantities as functions of $\protect%
\underline{\protect\pi }\equiv \protect\pi _{0}$ computed using the method
of heat potentials and the Monte Carlo method for $\overline{\protect\pi }=1$%
; $\protect\theta =1.0$, $T=1.96$.}
\label{fig:efg}
\end{figure}

\begin{figure}[h]
\begin{center}
\subfloat[]{\includegraphics[width=0.8\textwidth]
{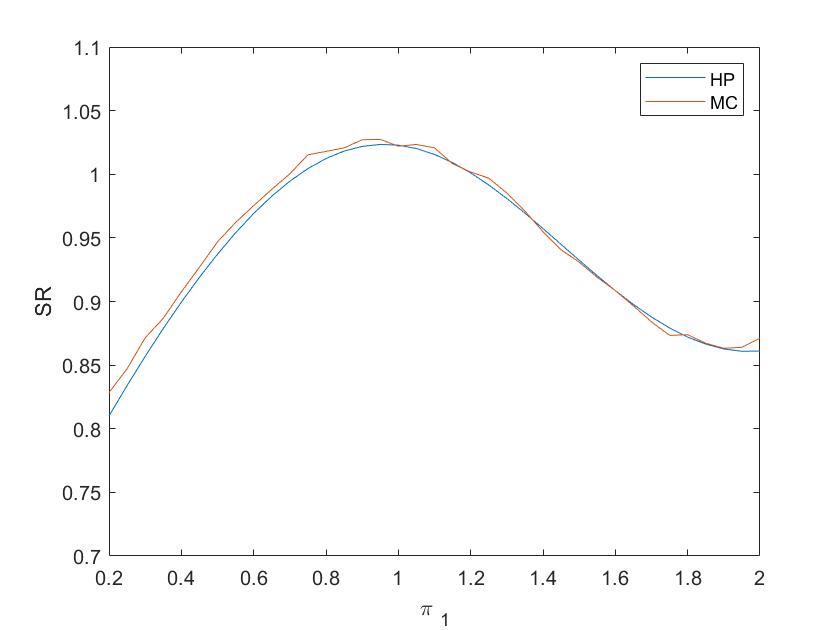}}
\\[0pt]
\subfloat[]{\includegraphics[width=0.8\textwidth]
{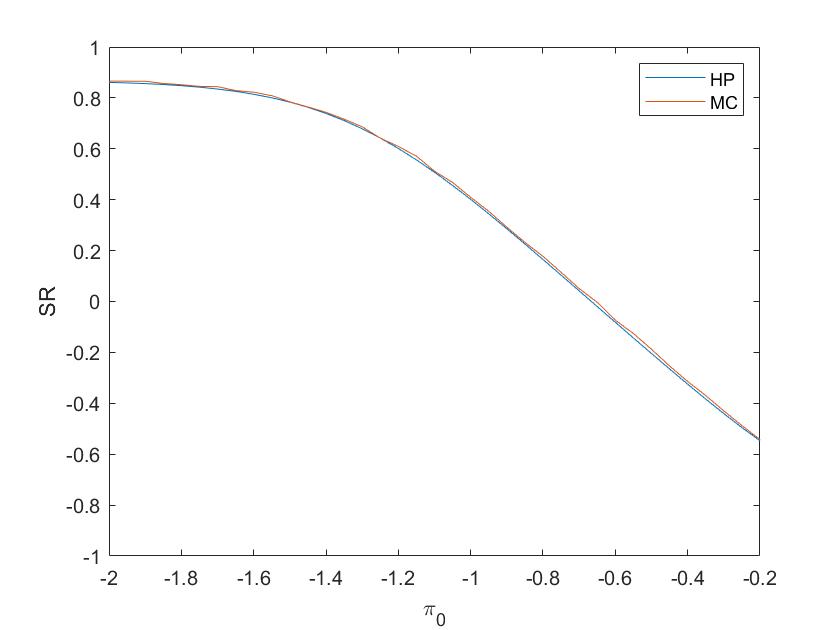}}
\end{center}
\par
\vspace{-10pt}
\caption{ (a) The Sharpe ratio as a function of $\overline{\protect\pi }%
\equiv \protect\pi _{1}$ computed by using the method of heat potentials and
the Monte Carlo method for $\protect\underline{\protect\pi }=-1$ (b) the
Sharpe ratio as a function of $\protect\underline{\protect\pi }\equiv 
\protect\pi _{0}$ computed using the method of heat potentials and the Monte
Carlo method for $\overline{\protect\pi }=1$; $\protect\theta =1.0$, $T=1.96$%
.}
\label{fig:sharpe}
\end{figure}

To be concrete, consider $\theta =1.0$ and $\Upsilon =0.49$, $T=1.96$. We
compare our results with the Monte Carlo method, which simulates the process
and compute its expectation and variance (see \cite{Lopez2018}). First, we
compare separately $E$, $\sigma =\sqrt{F-E^{2}}$, and $G$ calculated by both
methods in Figure \ref{fig:efg}:%
\begin{equation*}
\text{Figure \ref{fig:efg} near here.}
\end{equation*}%
Second, we show the results for the $\SR$ itself in Figure \ref{fig:sharpe}:

\begin{equation*}
\text{Figure \ref{fig:sharpe} near here.}
\end{equation*}%
We see that the relative difference between the method of heat potentials
and the Monte Carlo method is small and mainly comes from the Monte Carlo
noise.

\subsection{Optimization of the Sharpe ratio}

In this section we solve a problem of finding parameters to maximize the $%
\SR
$ by analyzing it as a function of $(\underline{\pi },\overline{\pi })$ for
different values of $\theta $ and $\Upsilon $. Two problems are considered:
(A) Fix $\Upsilon $ and maximize the $\SR$\ over $(\underline{\pi },%
\overline{\pi })$; (B) Maximize the $\SR$ over $(\underline{\pi },\overline{%
\pi },\Upsilon )$.

Given that the natural unit $\Omega =1/\sqrt{2}$, we consider three
representative values of $\theta $, namely $\theta =1$, $\theta =0.5$, and $%
\theta =0$, corresponding to strong and weak mispricing and fair pricing,
respectively. We choose three maturities, $\Upsilon =0.49$, $0.4999$, $%
0.499999$ or ,equivalently, $T=1.96$, $4.26$, $6.56$. For negative $\theta $%
, the corresponding $\SR$ can be obtained by reflection if needed.

We show the corresponding $\SR$ surfaces in Figures \ref{fig:theta=1.0}, \ref%
{fig:theta=0.5}, \ref{fig:theta=0.0}: 
\begin{equation*}
\text{Figure \ref{fig:theta=1.0} near here.}
\end{equation*}%
\begin{equation*}
\text{Figure \ref{fig:theta=0.5} near here.}
\end{equation*}%
\begin{equation*}
\text{Figure \ref{fig:theta=0.0} near here.}
\end{equation*}

The optimal bounds $(\underline{\pi }^{\ast },\overline{\pi }^{\ast })$ are
given in the Table \ref{Tab1} below:

\begin{equation*}
\text{Table \ref{Tab1} near here.}
\end{equation*}%
This table shows that in the case when the original mispricing is strong ($%
\theta =1$) it is not optimal to stop the trade early. When the mispricing
is weaker ($\theta =0.5$) or there is no mispricing in the first place ($%
\theta =0$) it is not optimal to stop losses, but it might be beneficial to
take profits. We emphasize that in practice one needs to use a highly
reliable estimation of the O-U parameters to be able to use these rules with
confidence.

\section{Traditional approaches}

\subsection{Motivation}

The method of heat potentials boils down to solving a system of Volterra
equations of the second kind. However, there are certain quantities of
interest, which can be calculated directly. To put it into a proper context,
in this section we discuss several classical approached to the problem we
are interested in. We emphasize that the method of heat potentials is
dramatically different from other method because it allows one to consider
strategies with finite duration, say $T$, whilst, to the best of our
knowledge, all other methods are asymptotic in nature and assume that $%
T\rightarrow \infty $.

\subsection{Expectation and variance of the trade's duration}

\label{Expectation}

In this subsection, we calculate the expected value and the variance of the
of duration of a trade, which terminates only when the spread hits one of
the barriers, $T=\infty $ (or $\Upsilon =0.5$). Specifically, we show how to
calculate these quantities analytically by solving inhomogeneous linear
ordinary differential equations (ODEs).

In the case in question, the second change of variables is not necessary, so
that we can concentrate on the following problems:%
\begin{equation}
\begin{array}{c}
G_{t}^{\left( 1\right) }\left( t,x\right) +\left( \theta -x\right)
G_{x}^{\left( 1\right) }\left( t,x\right) +\frac{1}{2}G_{xx}^{\left(
1\right) }\left( t,x\right) =0, \\ 
\\ 
G^{\left( 2\right) }\left( t,\underline{\pi }\right) =t,\ \ \ G^{\left(
1\right) }\left( t,\overline{\pi }\right) =t,%
\end{array}
\label{Eq16a}
\end{equation}%
\begin{equation}
\begin{array}{c}
G_{t}^{\left( 2\right) }\left( t,x\right) +\left( \theta -x\right)
G_{x}^{\left( 2\right) }\left( t,x\right) +\frac{1}{2}G_{xx}^{\left(
2\right) }\left( t,x\right) =0, \\ 
\\ 
G^{\left( 2\right) }\left( t,\underline{\pi }\right) =t^{2},\ \ \ G^{\left(
2\right) }\left( t,\overline{\pi }\right) =t^{2}.%
\end{array}
\label{Eq16}
\end{equation}%
with implicit terminal conditions at $T\rightarrow \infty $. The
superscripts indicate the first and second moments, respectively.

We start with the expectation. We can represent the solution $G^{\left(
1\right) }\left( t,x\right) $ of Eq. (\ref{Eq16a}) in a semi-stationary form:%
\begin{equation*}
\begin{array}{c}
G^{\left( 1\right) }\left( t,x\right) =t+g^{\left( 1\right) }\left( x\right)
,%
\end{array}%
\end{equation*}%
where%
\begin{equation}
\begin{array}{c}
\left( \theta -x\right) g_{x}^{\left( 1\right) }\left( x\right) +\frac{1}{2}%
g_{xx}^{\left( 1\right) }\left( x\right) =-1,%
\end{array}
\label{ODE2}
\end{equation}%
\begin{equation}
\begin{array}{c}
\ g^{\left( 1\right) }\left( \underline{\pi }\right) =0,\ \ \ \ g^{\left(
1\right) }\left( \overline{\pi }\right) =0.%
\end{array}
\label{BC}
\end{equation}

Eq. (\ref{ODE2}) can be solved by the method of variation of constants:%
\begin{equation}
\begin{array}{c}
g_{x}^{\left( 1\right) }\left( x\right) =a_{1}e^{\left( x-\theta \right)
^{2}}+\lambda \mathcal{F}\left( x-\theta \right) , \\ 
\\ 
g^{\left( 1\right) }\left( x\right) =a_{0}+a_{1}\mathcal{I}\left( x-\theta
\right) +\lambda \mathcal{G}\left( x-\theta \right) ,%
\end{array}
\label{Eq10}
\end{equation}%
where $a_{0}$, $a_{1}$ are arbitrary constants. Here $\mathcal{D}\left(
x\right) $ is Dawson's function, $\mathcal{E}\left( x\right) $ is its
integral, and $\mathcal{F}\left( x\right) ,\mathcal{G}\left( x\right) $ are
convenient abbreviations:%
\begin{equation*}
\begin{array}{c}
\mathcal{I}\left( x\right) =\int_{0}^{x}e^{z^{2}}dz,\ \ \ \mathcal{D}\left(
x\right) =e^{-x^{2}}\int_{0}^{x}e^{z^{2}}dz=e^{-x^{2}}\mathcal{I}\left(
x\right) , \\ 
\\ 
\mathcal{E}\left( x\right) =\int_{0}^{x}D\left( z\right) dz,\ \ \ \mathcal{F}%
\left( x\right) =\sqrt{\pi }N\left( \sqrt{2}x\right) e^{x^{2}}, \\ 
\\ 
\mathcal{G}\left( x\right) =\sqrt{\pi }N\left( \sqrt{2}x\right) \mathcal{I}%
\left( x\right) -\mathcal{E}\left( x\right) .%
\end{array}%
\end{equation*}%
We can use the Taylor series expansion for $\mathcal{G}\left( x\right) $ and
represent it in the form%
\begin{equation}
\begin{array}{c}
\mathcal{G}\left( x\right) =\frac{1}{4}\dsum\limits_{n=1}^{\infty }\frac{%
\Gamma \left( \frac{n}{2}\right) }{\Gamma \left( n+1\right) }\left(
2x\right) ^{n},%
\end{array}
\label{Eq14a}
\end{equation}%
see also \cite{Riccardi1988}, where this formula is obtained via the Laplace
transform. Taking into account boundary conditions (\ref{BC}), we can
represent $g$ as follows:%
\begin{equation}
\begin{array}{c}
g^{\left( 1\right) }\left( x,\underline{\pi },\overline{\pi }\right)
=2\left( \frac{\left( \mathcal{G}\left( \overline{\pi }-\theta \right) -%
\mathcal{G}\left( \underline{\pi }-\theta \right) \right) }{\left( \mathcal{I%
}\left( \overline{\pi }-\theta \right) -\mathcal{I}\left( \underline{\pi }%
-\theta \right) \right) }\left( \mathcal{I}\left( x-\theta \right) -\mathcal{%
I}\left( \underline{\pi }-\theta \right) \right) -\left( \mathcal{G}\left(
x-\theta \right) -\mathcal{G}\left( \underline{\pi }-\theta \right) \right)
\right) .%
\end{array}
\label{Eq14b}
\end{equation}%
Finally, the expected duration is given by the following expression:%
\begin{equation}
\begin{array}{c}
\DUR=g\left( 0\right) .%
\end{array}
\label{Dur}
\end{equation}

We show the expected duration as a function of $\underline{\pi },\overline{%
\pi }$ for $\theta =1$ in Figure \ref{fig:ed}: 
\begin{equation*}
\text{Figure \ref{fig:ed} near here.}
\end{equation*}%
Given the fact that $\Upsilon \rightarrow 0.5$ corresponds to $T\rightarrow
\infty $, we can see from this Figure that for sufficiently remote $%
\underline{\pi },\overline{\pi }$ the process stays within the range $\left[ 
\underline{\pi },\overline{\pi }\right] $ indefinitely, or, at least, for a
very long time.

Now we consider Eq. (\ref{Eq16}) and write%
\begin{equation*}
\begin{array}{c}
G^{\left( 2\right) }\left( t,x\right) =t^{2}+tg^{\left( 2,1\right) }\left(
x\right) +g^{\left( 2,0\right) }\left( x\right) ,%
\end{array}%
\end{equation*}%
where%
\begin{equation*}
\begin{array}{c}
\left( \theta -x\right) g_{x}^{\left( 2,1\right) }\left( x\right) +\frac{1}{2%
}g_{xx}^{\left( 2,1\right) }\left( x\right) =-2, \\ 
\\ 
g^{\left( 2,1\right) }\left( \underline{\pi }\right) =0,\ \ \ g^{\left(
2,1\right) }\left( \overline{\pi }\right) =0,%
\end{array}%
\end{equation*}%
\begin{equation}
\begin{array}{c}
\left( \theta -x\right) g_{x}^{\left( 2,0\right) }\left( x\right) +\frac{1}{2%
}g_{xx}^{\left( 2,0\right) }\left( x\right) =-g^{\left( 2,1\right) }\left(
x\right) , \\ 
\\ 
g^{\left( 2,0\right) }\left( \underline{\pi }\right) =0,\ \ \ g^{\left(
2,0\right) }\left( \overline{\pi }\right) =0.%
\end{array}
\label{Eq22}
\end{equation}%
In is clear that%
\begin{equation*}
\begin{array}{c}
g^{\left( 2,1\right) }\left( x\right) =2g^{\left( 1\right) }\left( x,%
\underline{\pi },\overline{\pi }\right) ,%
\end{array}%
\end{equation*}%
where $g^{\left( 1\right) }$ is given by Eq. (\ref{Eq14b}).

Green's function $\mathfrak{G}\left( x,y\right) $ for problem (\ref{Eq22})
has the form%
\begin{equation*}
\begin{array}{c}
\mathfrak{G}\left( x,y\right) =\left\{ 
\begin{array}{cc}
2\frac{e^{-\left( y-\theta \right) ^{2}}\left( \mathcal{I}\left( y-\theta
\right) -\mathcal{I}\left( \underline{\pi }-\theta \right) \right) }{\left( 
\mathcal{I}\left( \overline{\pi }-\theta \right) -\mathcal{I}\left( 
\underline{\pi }-\theta \right) \right) }\left( \mathcal{I}\left( x-\theta
\right) -\mathcal{I}\left( \overline{\pi }-\theta \right) \right) & y\leq
x\leq \overline{\pi }, \\ 
2\frac{e^{-\left( y-\theta \right) ^{2}}\left( \mathcal{I}\left( y-\theta
\right) -\mathcal{I}\left( \overline{\pi }-\theta \right) \right) }{\left( 
\mathcal{I}\left( \overline{\pi }-\theta \right) -\mathcal{I}\left( 
\underline{\pi }-\theta \right) \right) }\left( \mathcal{I}\left( x-\theta
\right) -\mathcal{I}\left( \underline{\pi }-\theta \right) \right) & 
\underline{\pi }\leq x\leq y.%
\end{array}%
\right.%
\end{array}%
\end{equation*}%
As usual,%
\begin{equation*}
\begin{array}{c}
g^{\left( 2,0\right) }\left( x\right) =-2\int_{-\infty }^{u}\mathfrak{G}%
\left( x,y\right) g^{\left( 1\right) }\left( y,\underline{\pi },\overline{%
\pi }\right) dy.%
\end{array}%
\end{equation*}%
The explicit expression for the expected duration given by Eq. (\ref{Dur})
is interesting in its own right and also can be used for benchmarking
solutions obtained via the method of heat potentials.

\subsection{Renewal theory approach}

To facilitate the comparison with previously know results, from now on, we
assume that $\theta =0$. 

In this subsection, we revisit Bertram's approach \cite{Bertram2009,
Bertram2010}. In a nutshell, Bertram assumes that the underlying O-U
process, representing portfolio's log-price, is running in perpetuity. He
envisions the following investment strategy. When the return process $x$
hits the lower level $l$, the underlying is bought. When the process $x$
hits the upper level $u$, the underlying is sold. Thus, the round trip is
characterized by two transitions, $x=l\rightarrow x=u$, and $x=u\rightarrow
x=l$; once the round trip is completed, the process starts again.

We can use the same ideas as in Section \ref{Expectation} to calculate $%
\mathbb{E}\left( T\right) $,$\mathbb{E}\left( T^{2}\right) $, and $\mathbb{V}%
\left( T\right) $ for the hitting time of a given level $u$, starting at the
level $x=l$ by letting $\underline{\pi }\rightarrow -\infty ,\overline{\pi }%
=u$: 
\begin{equation*}
\begin{array}{c}
\mathbb{E}\left( T\right) =2\left( \mathcal{G}\left( u\right) -\mathcal{G}%
\left( l\right) \right) , \\ 
\\ 
\mathbb{E}\left( T^{2}\right) =8\left( \mathcal{G}\left( u\right) \left( 
\mathcal{G}\left( u\right) -\mathcal{G}\left( l\right) \right) -\left( 
\mathcal{J}\left( u\right) -\mathcal{J}\left( l\right) \right) \right) , \\ 
\\ 
\mathbb{V}\left( T\right) =4\left( \left( \mathcal{G}^{2}\left( u\right) -2%
\mathcal{J}\left( u\right) \right) -\left( \mathcal{G}^{2}\left( l\right) -2%
\mathcal{J}\left( l\right) \right) \right) ,%
\end{array}%
\end{equation*}%
where%
\begin{equation*}
\begin{array}{c}
\mathcal{J}\left( x\right) =\int_{-\infty }^{x}e^{-y^{2}}\left( \mathcal{I}%
\left( x\right) -\mathcal{I}\left( y\right) \right) \mathcal{G}\left(
y\right) dy \\ 
\\ 
=\mathcal{I}\left( x\right) \int_{-\infty }^{x}e^{-y^{2}}\mathcal{G}\left(
y\right) dy-\int_{-\infty }^{x}\mathcal{D}\left( y\right) \mathcal{G}\left(
y\right) dy.%
\end{array}%
\end{equation*}%
Similarly to Eq. (\ref{Eq14a}), we can write:%
\begin{equation*}
\begin{array}{c}
\mathcal{J}\left( x\right) =\frac{1}{16}\dsum\limits_{n=1}^{\infty }\frac{%
\Gamma \left( \frac{n}{2}\right) \left( \psi \left( \frac{n}{2}\right)
+\gamma \right) }{\Gamma \left( n+1\right) }\left( 2x\right) ^{n},%
\end{array}%
\end{equation*}%
where $\psi $ is the digamma function and $\gamma $ is the Euler-Mascheroni
constant, $\gamma =-\psi \left( 1\right) $, see also \cite{Riccardi1988},
where this formula is obtained via the Laplace transform.

In summary,%
\begin{equation*}
\begin{array}{c}
\varepsilon \left( l\rightarrow u\right) \equiv \mathbb{E}\left( T\right)
=2\left( \mathcal{G}\left( u\right) -\mathcal{G}\left( l\right) \right) , \\ 
\\ 
\vartheta \left( l\rightarrow u\right) \equiv \mathbb{V}\left( T\right)
=4\left( \left( \mathcal{G}^{2}\left( u\right) -2\mathcal{J}\left( u\right)
\right) -\left( \mathcal{G}^{2}\left( l\right) -2\mathcal{J}\left( l\right)
\right) \right) .%
\end{array}%
\end{equation*}%
By symmetry,%
\begin{equation*}
\begin{array}{c}
\varepsilon \left( u\rightarrow l\right) =\varepsilon \left( -u\rightarrow
-l\right) =2\left( \mathcal{G}\left( -l\right) -\mathcal{G}\left( -u\right)
\right) , \\ 
\\ 
\vartheta \left( u\rightarrow l\right) =\vartheta \left( -u\rightarrow
-l\right) =4\left( \left( \mathcal{G}^{2}\left( -l\right) -2\mathcal{J}%
\left( -l\right) \right) -\left( \mathcal{G}^{2}\left( -u\right) -2\mathcal{J%
}\left( -u\right) \right) \right) .%
\end{array}%
\end{equation*}%
Finally,%
\begin{equation*}
\begin{array}{c}
\varepsilon \left( l\rightarrow u\rightarrow l\right) \equiv \varepsilon
\left( l\rightarrow u\right) +\varepsilon \left( u\rightarrow l\right) \\ 
\\ 
=2\left( \mathcal{G}\left( u\right) -\mathcal{G}\left( -u\right) -\left( 
\mathcal{G}\left( l\right) -\mathcal{G}\left( -l\right) \right) \right) \\ 
\\ 
=2\sqrt{\pi }\left( \mathcal{I}\left( u\right) -\mathcal{I}\left( l\right)
\right) ,%
\end{array}%
\end{equation*}%
\begin{equation*}
\begin{array}{c}
\vartheta \left( l\rightarrow u\rightarrow l\right) \equiv \vartheta \left(
l\rightarrow u\right) +\vartheta \left( u\rightarrow l\right) \\ 
\\ 
=4\left( \left( \mathcal{G}^{2}\left( u\right) -\mathcal{G}^{2}\left(
-u\right) -2\left( \mathcal{J}\left( u\right) -\mathcal{J}\left( -u\right)
\right) \right) -\left( \mathcal{G}^{2}\left( l\right) -\mathcal{G}%
^{2}\left( -l\right) -2\left( \mathcal{J}\left( l\right) -\mathcal{J}\left(
-l\right) \right) \right) \right) \\ 
\\ 
=16\left( \left( \mathcal{G}^{\left( e\right) }\left( u\right) \mathcal{G}%
^{\left( o\right) }\left( u\right) -\mathcal{J}^{\left( o\right) }\left(
u\right) \right) -\left( \mathcal{G}^{\left( e\right) }\left( l\right) 
\mathcal{G}^{\left( o\right) }\left( l\right) -\mathcal{J}^{\left( o\right)
}\left( l\right) \right) \right)%
\end{array}%
\end{equation*}%
since $\mathcal{G}$, $\mathcal{J}$ are decomposed into the even and odd
parts as follows:%
\begin{equation*}
\begin{array}{c}
\mathcal{G}\left( x\right) =\mathcal{G}^{\left( e\right) }\left( x\right) +%
\mathcal{G}^{\left( o\right) }\left( x\right) , \\ 
\\ 
\mathcal{G}^{\left( e\right) }\left( x\right) =\sqrt{\pi }\left( N\left( 
\sqrt{2}x\right) -\frac{1}{2}\right) \mathcal{I}\left( x\right) -\mathcal{E}%
\left( x\right) , \\ 
\\ 
\mathcal{G}^{\left( o\right) }\left( x\right) =\frac{\sqrt{\pi }}{2}\mathcal{%
I}\left( x\right) ,%
\end{array}%
\end{equation*}%
\begin{equation*}
\begin{array}{c}
\mathcal{J}\left( x\right) =\mathcal{J}^{\left( e\right) }\left( x\right) +%
\mathcal{J}^{\left( o\right) }\left( x\right) , \\ 
\\ 
\mathcal{J}^{\left( e\right) }\left( x\right) =\mathcal{I}\left( x\right)
\int_{0}^{x}e^{-y^{2}}\mathcal{G}^{\left( e\right) }\left( y\right)
dy-\int_{-\infty }^{0}\mathcal{D}\left( y\right) \mathcal{G}\left( y\right)
dy-\int_{0}^{x}\mathcal{D}\left( y\right) \mathcal{G}^{\left( e\right)
}\left( y\right) dy \\ 
\\ 
\mathcal{J}^{\left( o\right) }\left( x\right) =\mathcal{I}\left( x\right)
\left( \int_{-\infty }^{0}e^{-y^{2}}\mathcal{G}\left( y\right) dy+\frac{%
\sqrt{\pi }}{2}\mathcal{E}\left( x\right) \right) -\frac{\sqrt{\pi }}{2}%
\int_{0}^{x}\mathcal{D}\left( y\right) \mathcal{I}\left( y\right) dy.%
\end{array}%
\end{equation*}

Once the requisite quantities are computed, Bertram invokes classical
results from renewal theory, \cite{Bertram2009, Bertram2010}.\footnote{%
We note in passing that Bertram uses informal notation, which is
dimensionally incorrect, such as $\mathbb{V}\left( 1/T\right) =\mathbb{V}%
\left( T\right) /\mathbb{E}\left( T\right) ^{3}$, and next to impossible to
understand, although his final results are correct.} The classical result
from renewal theory, see, e.g., \cite{Ross2010}, gives the asymptotic
properties of the random variable $M\left( t,l,u\right) $ representing the
number of round trips on the time interval $\left[ 0,t\right] $:%
\begin{equation*}
\begin{array}{c}
M\left( t,l,u\right) \sim \mathfrak{N}\left( \frac{t}{\varepsilon \left(
l\rightarrow u\rightarrow l\right) },\frac{\vartheta \left( l\rightarrow
u\rightarrow l\right) t}{\varepsilon \left( l\rightarrow u\rightarrow
l\right) ^{3}}\right) ,%
\end{array}%
\end{equation*}%
where $\mathfrak{N}$ is the normal variable. Given that $x$ represents the
log-price of the underlying portfolio, the return and asymptotic Sharpe
ratio $\SR$ per unit of time are given by%
\begin{equation*}
\begin{array}{c}
r=\frac{\left( u-l-c\right) }{\varepsilon \left( l\rightarrow u\rightarrow
l\right) }, \\ 
\\ 
\SR=\sqrt{\frac{\varepsilon \left( l\rightarrow u\rightarrow l\right) }{%
\vartheta \left( l\rightarrow u\rightarrow l\right) }}\frac{\left(
u-l-c-r_{f}\right) }{\left( u-l-c\right) },%
\end{array}%
\end{equation*}%
where $c$, $r_{f}$ represent transaction fees, and risk-free rate,
respectively. Bertram maximizes one of these quantities over the
stop-loss/take profit thresholds $\left( l,u\right) $.

The main practical problem with this approach is that it assumes
stationarity in perpetuity of the underlying process, which is a somewhat
questionable assumption. The other problem is that, even for a stationary
process, it takes a very long time for the strategy to reach its asymptotic
state. The reason why these issues have not been discussed earlier, is that
it is very hard to calculate the probability density function (pdf) for the
processes $l\rightarrow u$, $u\rightarrow l$, and the round-trip process $%
l\rightarrow u\rightarrow l$. Recently, Lipton and Kaushansky proposed a
very efficient method for calculating the pdfs for the processes $%
l\rightarrow u$, $u\rightarrow l$, see \cite{Lipton2018, Lipton2020a}; the
the round-trip process $l\rightarrow u\rightarrow l$ can be analyzed by
convolution. We show the corresponding pdfs for a representative choice of $l
$, $u$, namely $l=-1/\sqrt{2}$, $u=1/\sqrt{2}$ in Figure \ref{Fig12}. This
figure clearly shows that a very long right tail characterizes the
round-trip process $l\rightarrow u\rightarrow l$, so that the strategy might
never reach its asymptotic limit in practice.%
\begin{equation*}
\text{Figure \ref{Fig12} near here.}
\end{equation*}

\subsection{Perpetual value function}

In this section, we discuss results obtained in \cite{Ekstrom2010,
Ekstrom2011, Larsson2013, Lindberg2014}, and rederive and improve their
findings in a concise semi-analytic fashion.

The stationary problem for determining the value function and the \emph{%
optimal} take-profit level $u$ for a \emph{given} stop-loss level $l$ (which
is determined by the investor's risk appetite) and the time value of money
has the form:%
\begin{equation}
\begin{array}{c}
V_{xx}\left( x\right) -2xV_{x}\left( x\right) -\lambda V=0,\ \ \ l\leq x\leq
u, \\ 
\\ 
V\left( l\right) =l,\ \ \ V\left( u\right) =u,\ \ \ V^{\prime }\left(
u\right) =1.%
\end{array}
\label{Eq1}
\end{equation}%
This problem is similar, but by no means identical, to the pricing problem
for the perpetual American call option on a dividend-paying stock. Here $%
\lambda $ is the non-dimensional discount rate, $\lambda =2r/\kappa $.

The second-order ordinary differential equation (\ref{Eq1}) is the
well-known Hermite differential equation. Its general solution has the form%
\begin{equation}
\begin{array}{c}
V\left( x\right) =a_{0}M\left( \frac{\lambda }{4},\frac{1}{2},x^{2}\right)
+a_{1}xM\left( \frac{\lambda +2}{4},\frac{3}{2},x^{2}\right) ,%
\end{array}
\label{Eq4}
\end{equation}%
where $M\left( a,b,z\right) $ is the celebrated Kummer function (a confluent
hypergeometric function of the first kind), and $a_{0},a_{1}$ are arbitrary
constants. Boundary conditions (\ref{Eq1}) yield:%
\begin{equation*}
\begin{array}{c}
a_{0}M\left( \frac{\lambda }{4},\frac{1}{2},l^{2}\right) +a_{1}lM\left( 
\frac{\lambda +2}{4},\frac{3}{2},l^{2}\right) =l, \\ 
\\ 
a_{0}M\left( \frac{\lambda }{4},\frac{1}{2},u^{2}\right) +a_{1}uM\left( 
\frac{\lambda +2}{4},\frac{3}{2},u^{2}\right) =u, \\ 
\\ 
a_{0}\lambda uM\left( \frac{\lambda +4}{4},\frac{3}{2},u^{2}\right)
+a_{1}\left( M\left( \frac{\lambda +2}{4},\frac{3}{2},u^{2}\right) +\frac{%
\left( \lambda +2\right) }{3}u^{2}M\left( \frac{\lambda +6}{4},\frac{5}{2}%
,u^{2}\right) \right) =1.%
\end{array}%
\end{equation*}%
Here we use the fact that%
\begin{equation*}
\begin{array}{c}
M_{z}\left( a,b,z\right) =\frac{a}{b}M\left( a+1,b+1,z\right) .%
\end{array}%
\end{equation*}%
We eliminate $a_{0},a_{1}$:%
\begin{equation}
\begin{array}{c}
a_{0}=\frac{c_{11}l-c_{01}u}{c_{00}c_{11}-c_{01}c_{10}},\ \ \ a_{1}=\frac{%
-c_{10}l+c_{00}u}{c_{00}c_{11}-c_{01}c_{10}}, \\ 
\\ 
c_{00}=M\left( \frac{\lambda }{4},\frac{1}{2},l^{2}\right) ,\ \ \
c_{01}=lM\left( \frac{\lambda +2}{4},\frac{3}{2},l^{2}\right) , \\ 
\\ 
c_{10}=M\left( \frac{\lambda }{4},\frac{1}{2},u^{2}\right) ,\ \ \
c_{11}=uM\left( \frac{\lambda +2}{4},\frac{3}{2},u^{2}\right) ,%
\end{array}
\label{Eq7}
\end{equation}%
and obtain the following nonlinear algebraic equation for $u$:%
\begin{equation}
\begin{array}{c}
\left( c_{11}l-c_{01}u\right) \lambda uM\left( \frac{\lambda +4}{4},\frac{3}{%
2},u^{2}\right) \\ 
\\ 
+\left( -c_{10}l+c_{00}u\right) \left( M\left( \frac{\lambda +2}{4},\frac{3}{%
2},u^{2}\right) +\frac{\left( \lambda +2\right) }{3}u^{2}M\left( \frac{%
\lambda +6}{4},\frac{5}{2},u^{2}\right) \right) =\left(
c_{00}c_{11}-c_{01}c_{10}\right) .%
\end{array}
\label{Eq8}
\end{equation}%
We solve Eq. (\ref{Eq8}) via the Newton-Raphson method.

Once $u$ is found, we use Eqs (\ref{Eq4}), (\ref{Eq7}), to construct the
value functions $V\left( x\right) $. We show $V\left( x\right) -x$ and $%
u\left( l\right) $ for several representative values of $\lambda $ in
Figures \ref{Fig7} (a), (b). 
\begin{equation*}
\text{Figure \ref{Fig7} near here.}
\end{equation*}

The stationary problem for determining the value function and the \emph{%
optimal} take-profit level $U$ for a \emph{given} stop-loss level $L$ (which
is determined by the investor's risk appetite) and the opportunity cost $c$
has the nondimensional form:%
\begin{equation}
\begin{array}{c}
V_{xx}\left( x\right) -2xV_{x}\left( x\right) =\lambda ,\ \ \ l\leq x\leq u,
\\ 
\\ 
V\left( l\right) =l,\ \ \ V\left( u\right) =u,\ \ \ V_{x}\left( u\right) =1,%
\end{array}
\label{Eq9}
\end{equation}%
where $\lambda $ is the non-dimensional opportunity cost, $\lambda
=2c/\kappa $. It is easy to show that the general solution of Eq. (\ref{Eq9}%
) has the form given by Eq. (\ref{Eq10}). As before we get the following set
of equations:%
\begin{equation*}
\begin{array}{c}
a_{0}+a_{1}\mathcal{I}\left( l\right) =l-\lambda \mathcal{G}\left( l\right) ,
\\ 
\\ 
a_{0}+a_{1}\mathcal{I}\left( u\right) =u-\lambda \mathcal{G}\left( u\right) ,
\\ 
\\ 
a_{1}e^{u^{2}}=1-\lambda \mathcal{F}\left( u\right) .%
\end{array}%
\end{equation*}%
Accordingly,%
\begin{equation*}
\begin{array}{c}
a_{0}=\frac{\mathcal{I}\left( u\right) \left( l-\lambda \mathcal{G}\left(
l\right) \right) -\mathcal{I}\left( l\right) \left( u-\lambda \mathcal{G}%
\left( u\right) \right) }{\left( \mathcal{I}\left( u\right) -\mathcal{I}%
\left( l\right) \right) }, \\ 
\\ 
a_{1}=\frac{u-\lambda \mathcal{G}\left( u\right) -\left( l-\lambda \mathcal{G%
}\left( l\right) \right) }{\left( \mathcal{I}\left( u\right) -\mathcal{I}%
\left( l\right) \right) }, \\ 
\\ 
\frac{e^{u^{2}}\left( u-\lambda \mathcal{G}\left( u\right) -\left( l-\lambda 
\mathcal{G}\left( l\right) \right) \right) }{\left( \mathcal{I}\left(
u\right) -\mathcal{I}\left( l\right) \right) }=1-\lambda \mathcal{F}\left(
u\right) .%
\end{array}%
\end{equation*}%
In Figure \ref{Fig8}(a), we show $V\left( x\right) -x$ for $l=-2.0$ for
several representative values of $\lambda $, the corresponding optimal
values of $u$ are $1.07$, $0.74$, $0.50$. In Figure \ref{Fig8}(b) we show
the optimal boundary $u\left( l\right) $.%
\begin{equation*}
\text{Figure \ref{Fig8} near here.}
\end{equation*}

It is natural to ask what happens when the underlying mean-reverting process
has a jump component, so that%
\begin{equation*}
\begin{array}{c}
dx=-xdt+dW_{t}+JdP_{t},%
\end{array}%
\end{equation*}%
where $P_{t}$ is a Poisson process with intensity $\nu $, and $J$ is the
jump magnitude, which is assumed to be a random variable with density
function $\phi \left( J\right) $, see \cite{Larsson2013}. Larsson \textit{et
al.} use the finite element method to solve the corresponding free boundary
problem. However, if $J$ has a double exponential distribution density
function $\phi \left( J\right) $,%
\begin{equation*}
\begin{array}{c}
\phi \left( J\right) =\kappa e^{-\kappa \left\vert J\right\vert },%
\end{array}%
\end{equation*}%
or, more generally, a hyper-exponential distribution, see, e.g., \cite%
{Lipton2002}, the problem can be solved in a semi-analytical fashion.

To this end, it is convenient to write the problem with jumps in terms of $%
v\left( x\right) =V\left( x\right) -x$: 
\begin{equation*}
\begin{array}{c}
v^{\prime \prime }\left( x\right) -2xv^{\prime }\left( x\right) +\omega
\left( I_{+}\left( x\right) +I_{-}\left( x\right) -v\left( x\right) \right)
=\lambda +\left( 2+\omega \right) x, \\ 
\\ 
I_{+}\left( x\right) =\dint\limits_{0}^{x-l}v\left( x-z\right) e^{-\kappa
z}dz=\dint\limits_{l}^{x}v\left( z\right) e^{-\kappa \left( x-z\right) }dz,
\\ 
\\ 
I_{-}\left( x\right) =\dint\limits_{0}^{u-x}v\left( x+z\right) e^{-\kappa
z}dz=\dint\limits_{x}^{u}v\left( z\right) e^{-\kappa \left( z-x\right) }dz,
\\ 
\\ 
v\left( l\right) =v\left( u\right) =v^{\prime }\left( u\right) =I_{+}\left(
l\right) =I_{-}\left( u\right) =0,%
\end{array}%
\end{equation*}%
where $\omega =\kappa \nu $. It can be written as an inhomogeneous system of
linear ODEs:%
\begin{equation*}
\begin{array}{c}
v^{\prime }\left( x\right) -w\left( x\right) =0 \\ 
\\ 
w^{\prime }\left( x\right) -2xw\left( x\right) +\omega \left( I_{+}\left(
x\right) +I_{-}\left( x\right) -v\left( x\right) \right) =\lambda +\left(
2+\omega \right) x, \\ 
\\ 
I_{+}^{\prime }\left( x\right) -v\left( x\right) +\kappa I_{+}\left(
x\right) =0, \\ 
\\ 
I_{-}\left( x\right) +v\left( x\right) -\kappa I_{-}\left( x\right) =0, \\ 
\\ 
v\left( l\right) =0,\ \ \ w\left( l\right) =c,\ \ \ I_{+}\left( l\right)
=0,\ \ \ I_{-}\left( l\right) =d.%
\end{array}%
\end{equation*}

This system can be solved by the method of shooting by choosing initial
values $c,d$ and the right endpoint of the computational interval $b$ to
satisfy the remaining boundary conditions%
\begin{equation}
\begin{array}{c}
\ v\left( u\right) =w\left( u\right) =I_{-}\left( u\right) =0.%
\end{array}
\label{Eq33}
\end{equation}%
or in the matrix form:%
\begin{equation*}
\begin{array}{c}
\left( 
\begin{array}{c}
v \\ 
w \\ 
I_{+} \\ 
I_{-}%
\end{array}%
\right) ^{\prime }+\left( 
\begin{array}{cccc}
0 & -1 & 0 & 0 \\ 
-\omega & -2x & \omega & \omega \\ 
-1 & 0 & \kappa & 0 \\ 
1 & 0 & 0 & -\kappa%
\end{array}%
\right) \left( 
\begin{array}{c}
v \\ 
w \\ 
I_{+} \\ 
I_{-}%
\end{array}%
\right) =\left( 
\begin{array}{c}
0 \\ 
\lambda +\left( 2+\omega \right) x \\ 
0 \\ 
0%
\end{array}%
\right) .%
\end{array}%
\end{equation*}

In Figure \ref{Fig9}(a), we show the solution vector $\left( v\left(
x\right) ,w\left( x\right) ,I_{+}\left( x\right) ,I_{-}\left( x\right)
\right) $ corresponding to the suboptimal choice of $u$. The shooting
parameters, $c,d$, are chosen in such a way, that two of the three boundary
conditions (\ref{Eq33}) are satisfied, $v\left( u\right) =I_{-}\left(
u\right) =0$. In Figure \ref{Fig9}(b), we show what happens when the upper
limit $u$ is chosen optimally, by using the Newton-Raphson method. For $%
u=1.18$ all three conditions (\ref{Eq33}) are met. In Figure \ref{Fig9}(c),
we demonstrate the quality of our numerical method by putting $\omega =0$
and comparing the corresponding numerical solution with the analytical
solution given by Eq. (\ref{Eq10}). The figure shows that the agreement is
excellent.%
\begin{equation*}
\text{Figure \ref{Fig9} near here.}
\end{equation*}

In Figure \ref{Fig10}(a), we show $v\left( x\right) $ for $l=-2.0$, the
corresponding optimal values of $u$ are $1.18$, $0.80$, $0.55$; we show $%
u\left( l\right) $ for several representative values of $\lambda $ in Figure %
\ref{Fig10}(b), while 
\begin{equation*}
\text{Figure \ref{Fig10} near here.}
\end{equation*}

\subsection{ Linear transaction costs}

Several researchers, including de Lataillade \textit{et al.}, \cite%
{Lataillade2012}, concentrated on the critical question on how linear
transaction costs affect the profitability of mean-reverting trading
strategies. An alternative treatment is given by \cite{Martin2011}, see also 
\cite{Do2012}. Denuded of all amenities, the approach of de Lataillade 
\textit{et al.} is almost identical to the method used by Hyer et al., \cite%
{Hyer1997}, for studying passport options.

de Lataillade \textit{et al.} reduce the problem to solving the following
Fredholm integral equation of the second kind%
\begin{equation}
\begin{array}{c}
g\left( x\right) -\int_{-q}^{q}K\left( x,y\right) g\left( y\right)
dy=f\left( x\right) ,%
\end{array}
\label{Eq35}
\end{equation}%
where%
\begin{equation*}
\begin{array}{c}
K\left( x,y\right) =\frac{\Theta \exp \left( -\frac{\left( \Theta y-x\right)
^{2}}{\left( \Theta ^{2}-1\right) }\right) }{\sqrt{\pi \left( \Theta
^{2}-1\right) }}, \\ 
f\left( x\right) =x+\Gamma \left( N\left( -\frac{\sqrt{2}\left( \Theta
q-x\right) }{\sqrt{\left( \Theta ^{2}-1\right) }}\right) -N\left( -\frac{%
\sqrt{2}\left( \Theta q+x\right) }{\sqrt{\left( \Theta ^{2}-1\right) }}%
\right) \right) ,%
\end{array}%
\end{equation*}%
where $\Theta =e^{\Delta }$. Eq. (\ref{Eq35}) is augmented with the matching
condition%
\begin{equation}
\begin{array}{c}
g\left( q\right) =\Gamma .%
\end{array}
\label{Eq37}
\end{equation}%
Here $\Gamma $ represents transaction cost, while $\Delta $ shows how far
forward the behavior of the process can be predicted. The trader should not
change her position when $-q<x<q$, and go maximally long when $x=q$, and
short when $x=-q$.

While de Lataillade \textit{et al.} use the path integral method to
understand the behavior of Eqs (\ref{Eq35},\ref{Eq37}), we prefer to attack
the problem in question directly - by solving the corresponding Fredholm
equation. As before, we solve Eq. (\ref{Eq35}) for a given $q$, and then
adjust $q$ by using the Newton-Raphson method until the matching condition (%
\ref{Eq37}) is met. We notice in passing that $K\left( x,y\right) $ is even, 
$K\left( -x,-y\right) =K\left( x,y\right) $, while $f\left( x\right) $ is
odd, $f\left( -x\right) =-f\left( x\right) $, so that $g\left( x\right) $ is
even, $g\left( -x\right) =-g\left( x\right) $. Our analysis results in some
unexpected findings. Namely, Eqs (\ref{Eq35},\ref{Eq37}) have multiple
solutions. We choose $\Gamma =0.1$, $\Delta =1$ and solve the equations in
question. It turns out that at least two critical values of $q$ are
possible, $q=0.0561$ and $q=1.0131$. We show the corresponding solutions in
Figures \ref{Fig11}(a), (b). It can be shown that $g\left( x\right) $, which
has a single root at $x=0$ is the solution of interest. With this in mind,
we can construct critical boundaries $q\left( \Delta \right) $ corresponding
to several representative values of $\Gamma $. These boundaries are shown in
Figure \ref{Fig11}(c). 
\begin{equation*}
\text{Figure \ref{Fig11} near here.}
\end{equation*}

\section{Conclusions}

In this paper we create an analytical framework for computing optimal
stop-loss/take-profit bounds $(\underline{\pi }^{\ast },\overline{\pi }%
^{\ast })$ for O-U driven trading strategies by using the method of heat
potentials.

\begin{figure}[tbp]
\begin{center}
\subfloat[]{\includegraphics[width=0.5\textwidth]
{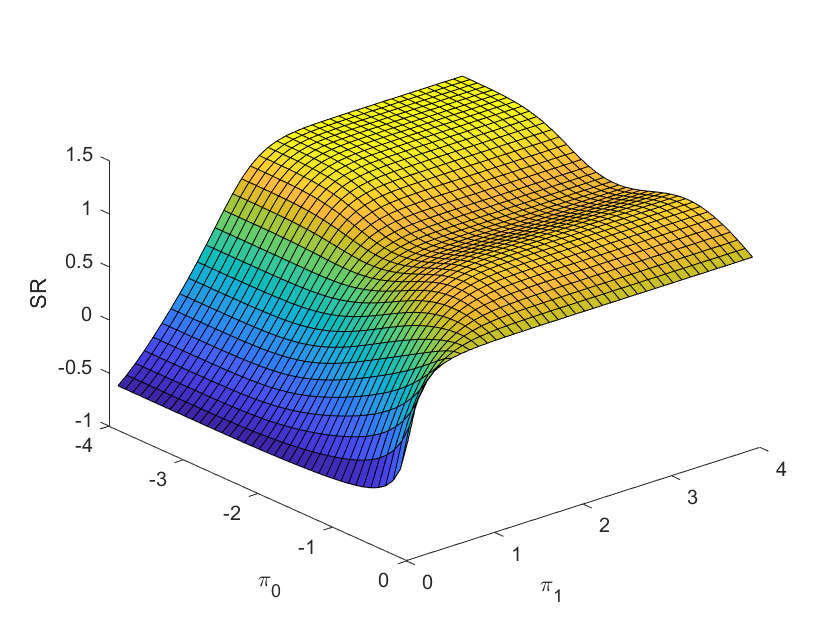}}
\subfloat[]{\includegraphics[width=0.5\textwidth]
{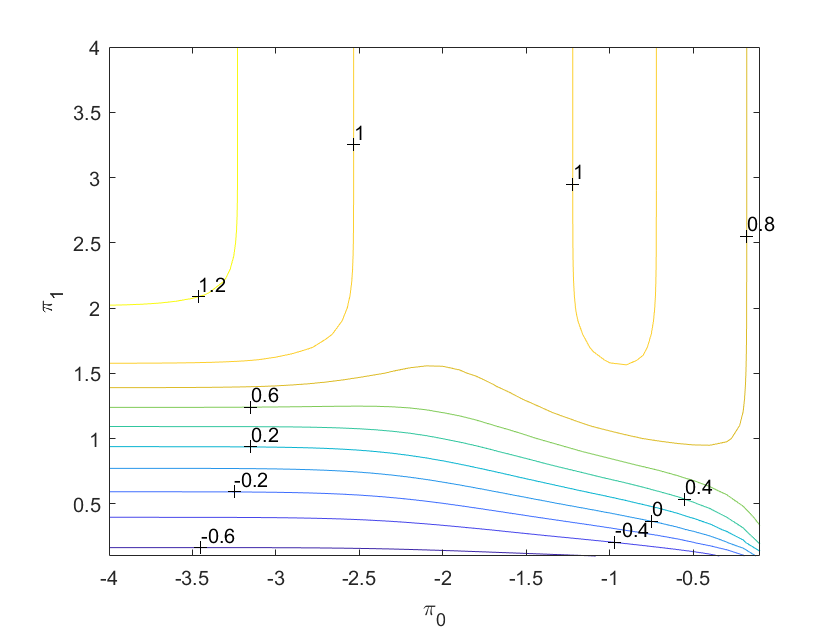}}%
\\[0pt]
\subfloat[]{\includegraphics[width=0.5\textwidth]
{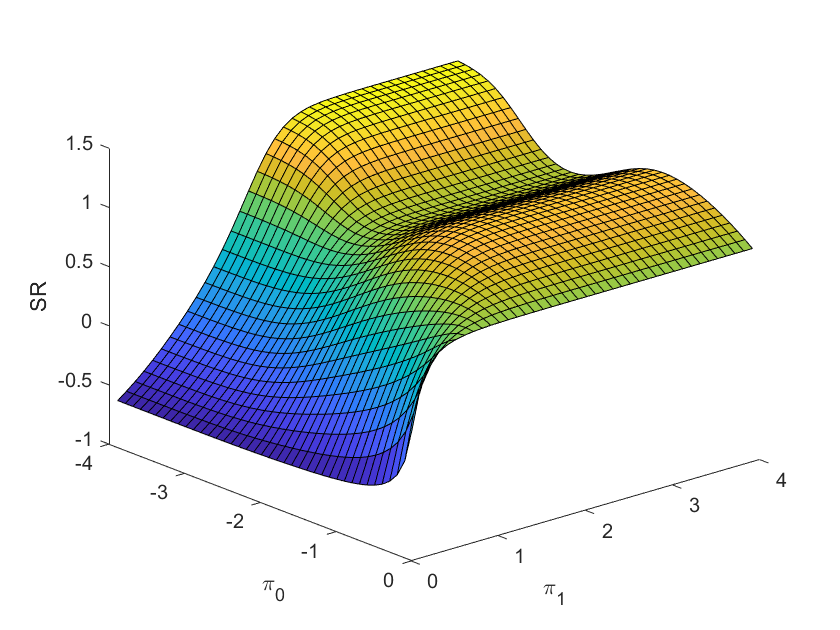}}
\subfloat[]{\includegraphics[width=0.5\textwidth]
{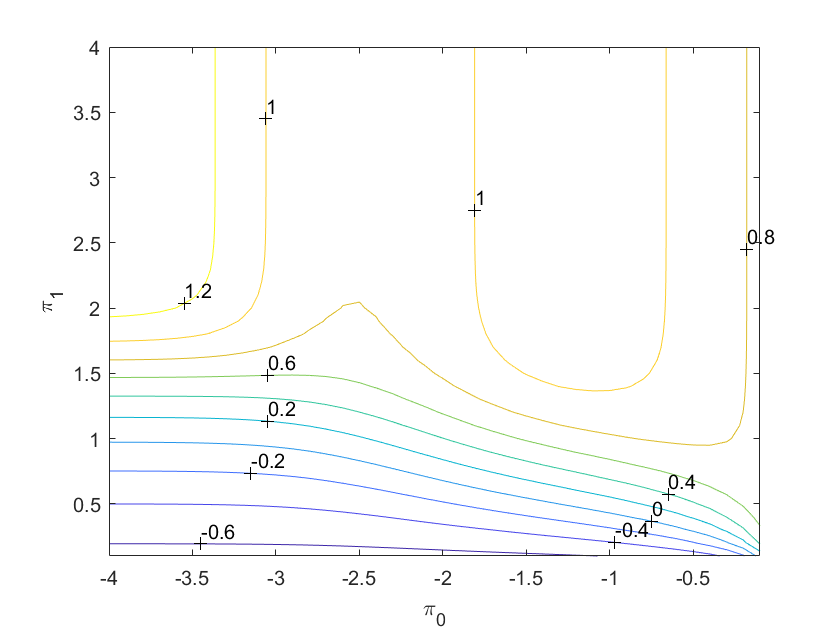}}%
\\[0pt]
\subfloat[]{\includegraphics[width=0.5\textwidth]
{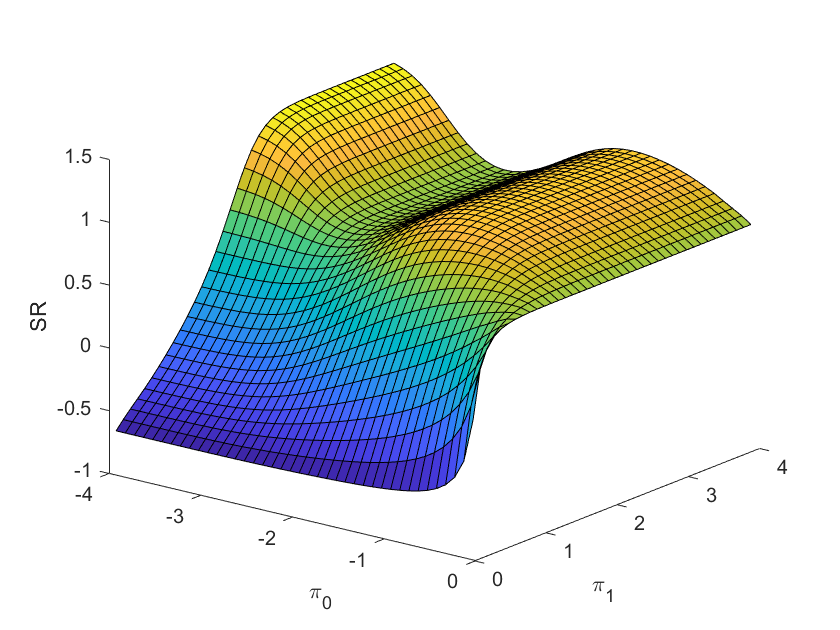}}
\subfloat[]{\includegraphics[width=0.5\textwidth]
{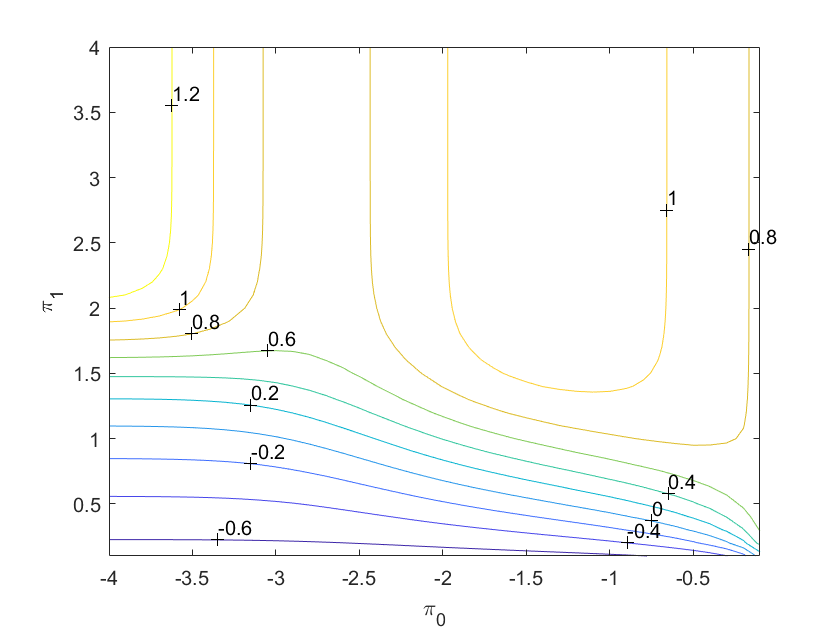}}
\end{center}
\par
\vspace{-10pt}
\caption{ The Sharpe Ratio as a function of $(\protect\underline{\protect\pi 
},\overline{\protect\pi })$ for $\protect\theta =1.0$ a)-b) $T=1.96$, c)-d) $%
T=4.26$, e)-f) $T=6.56$. }
\label{fig:theta=1.0}
\end{figure}

\begin{figure}[tbp]
\begin{center}
\subfloat[]{\includegraphics[width=0.5\textwidth]
{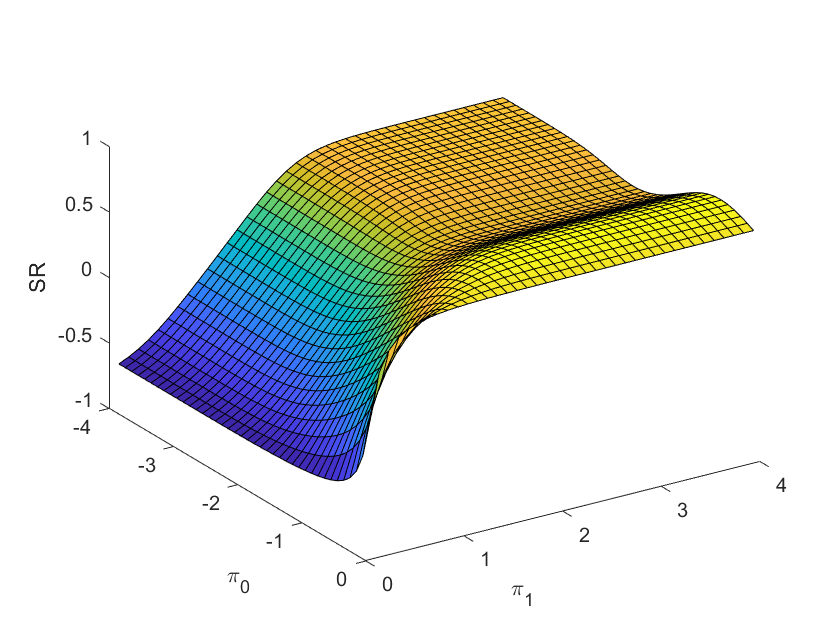}}
\subfloat[]{\includegraphics[width=0.5\textwidth]
{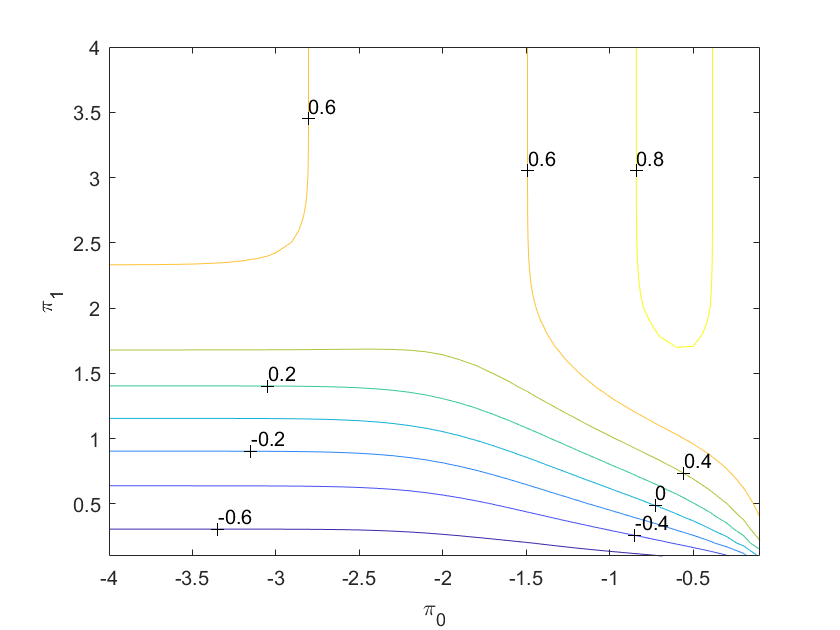}}%
\\[0pt]
\subfloat[]{\includegraphics[width=0.5\textwidth]
{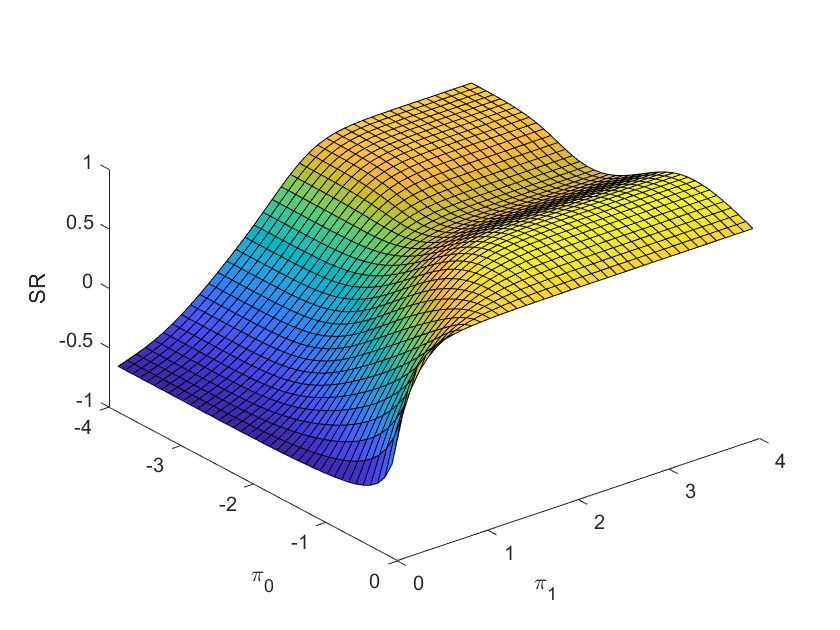}}
\subfloat[]{\includegraphics[width=0.5\textwidth]
{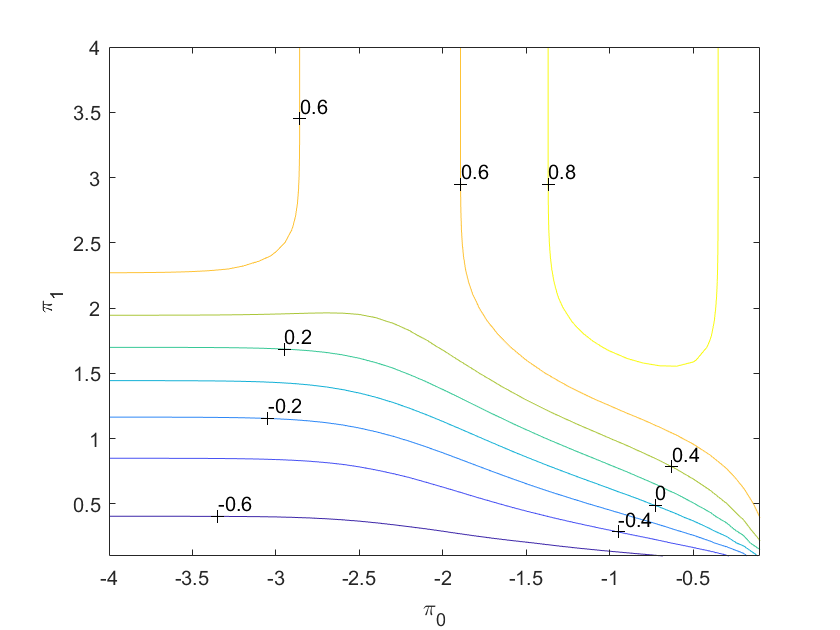}}%
\\[0pt]
\subfloat[]{\includegraphics[width=0.5\textwidth]
{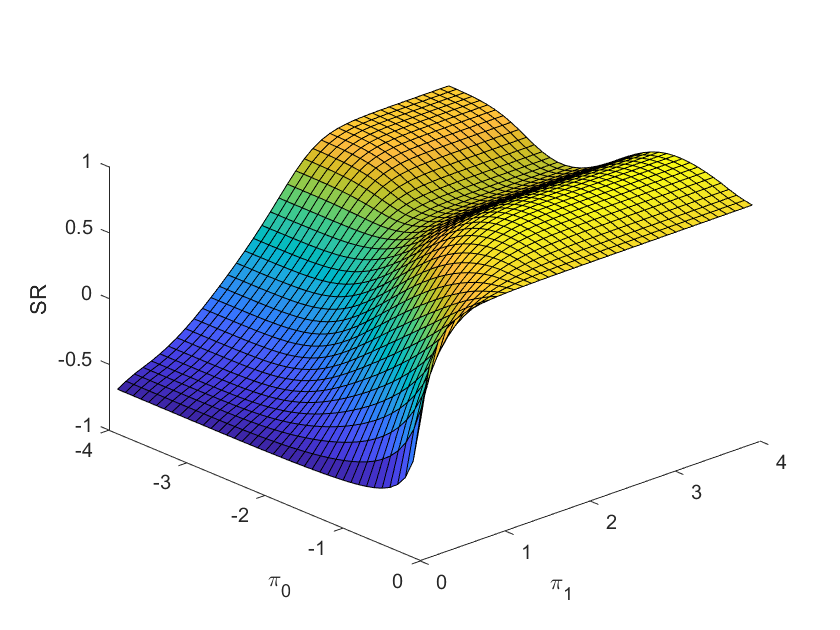}}
\subfloat[]{\includegraphics[width=0.5\textwidth]
{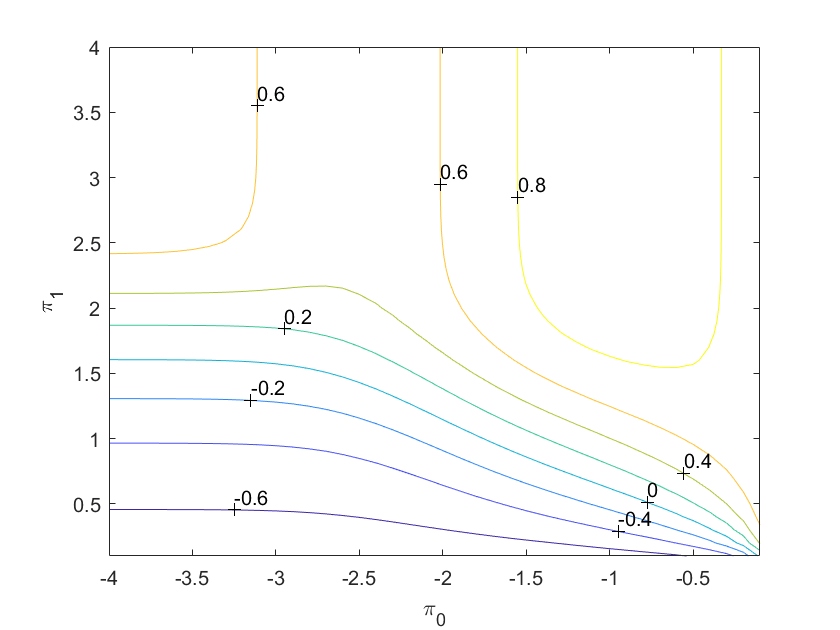}}
\end{center}
\par
\vspace{-10pt}
\caption{ The Sharpe Ratio as a function of $(\protect\underline{\protect\pi 
},\overline{\protect\pi })$ for $\protect\theta =0.5$ a)-b) $T=1.96$, c)-d) $%
T=4.26$, e)-f) $T=6.56$.}
\label{fig:theta=0.5}
\end{figure}

\begin{figure}[tbp]
\begin{center}
\subfloat[]{\includegraphics[width=0.5\textwidth]
{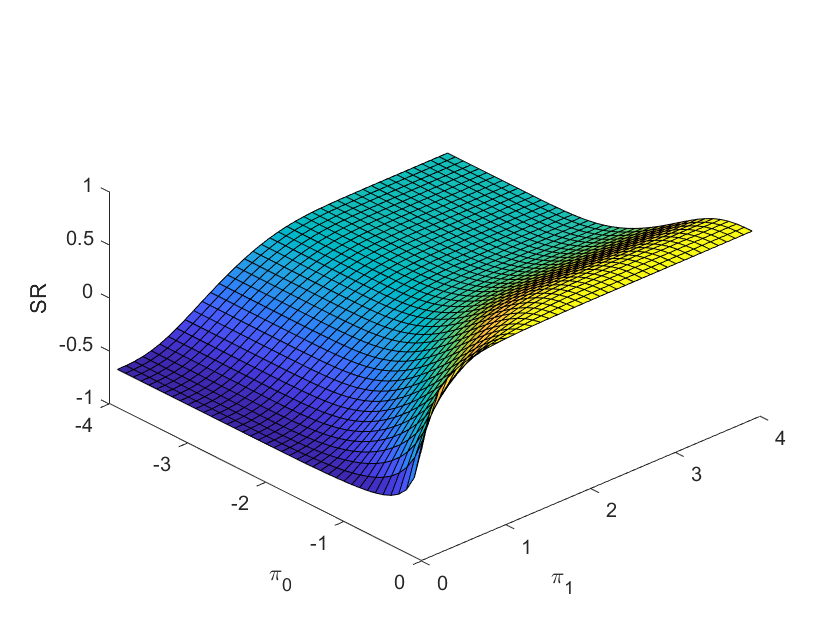}}
\subfloat[]{\includegraphics[width=0.5\textwidth]
{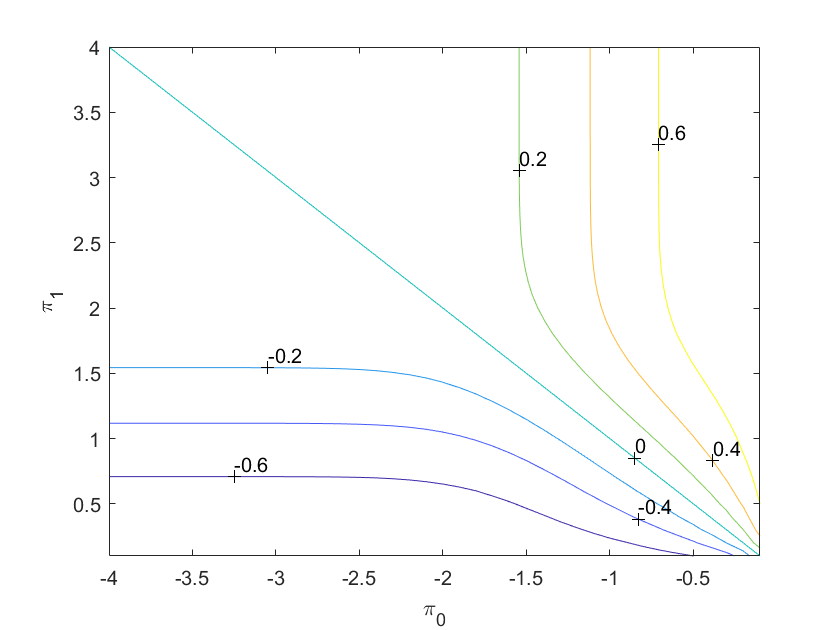}}%
\\[0pt]
\subfloat[]{\includegraphics[width=0.5\textwidth]
{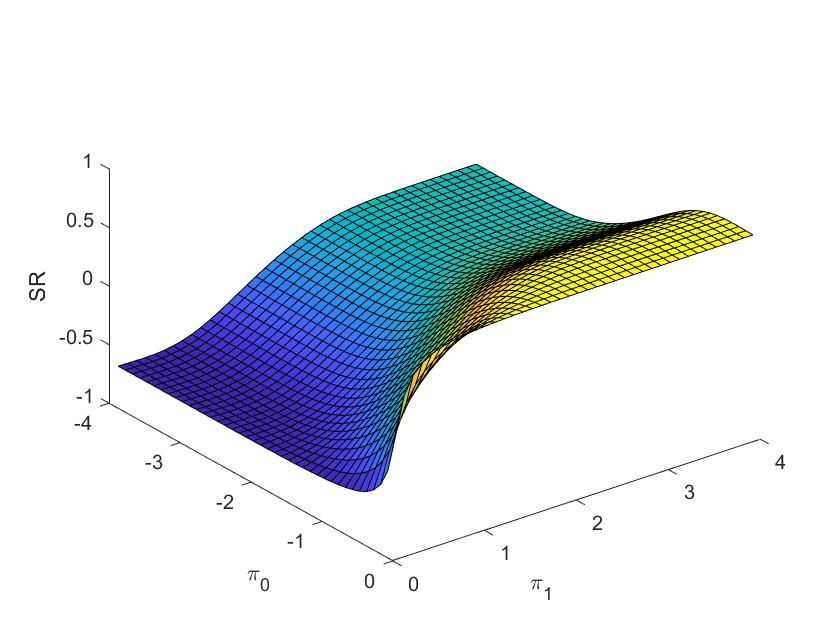}}
\subfloat[]{\includegraphics[width=0.5\textwidth]
{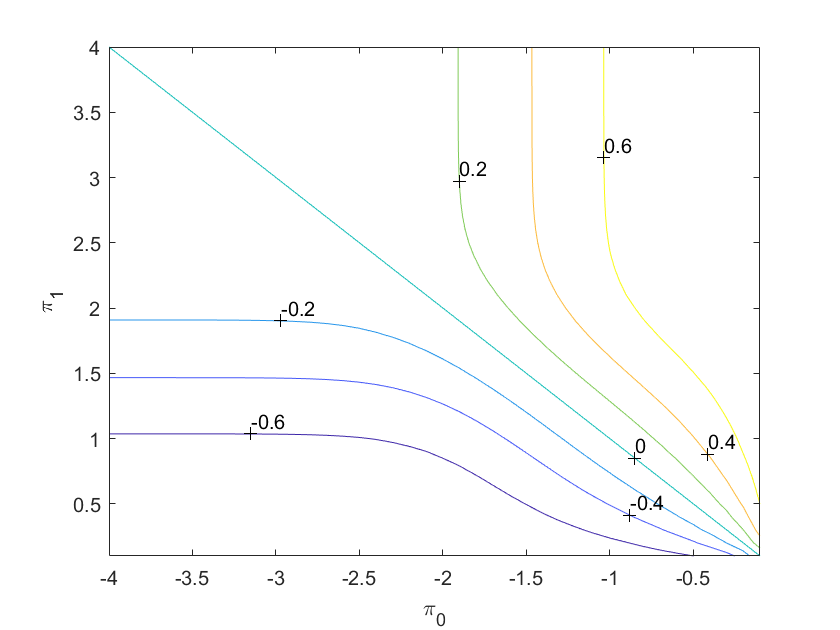}}%
\\[0pt]
\subfloat[]{\includegraphics[width=0.5\textwidth]
{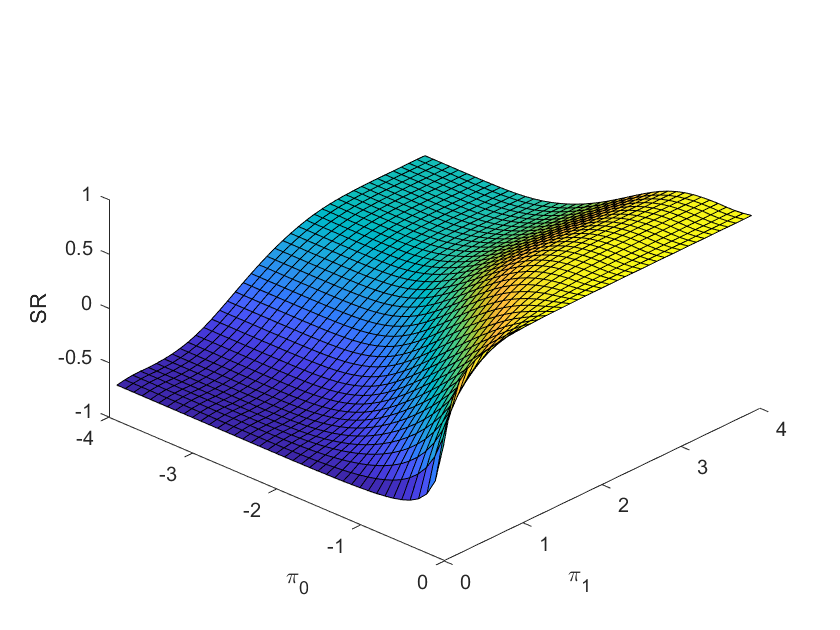}}
\subfloat[]{\includegraphics[width=0.5\textwidth]
{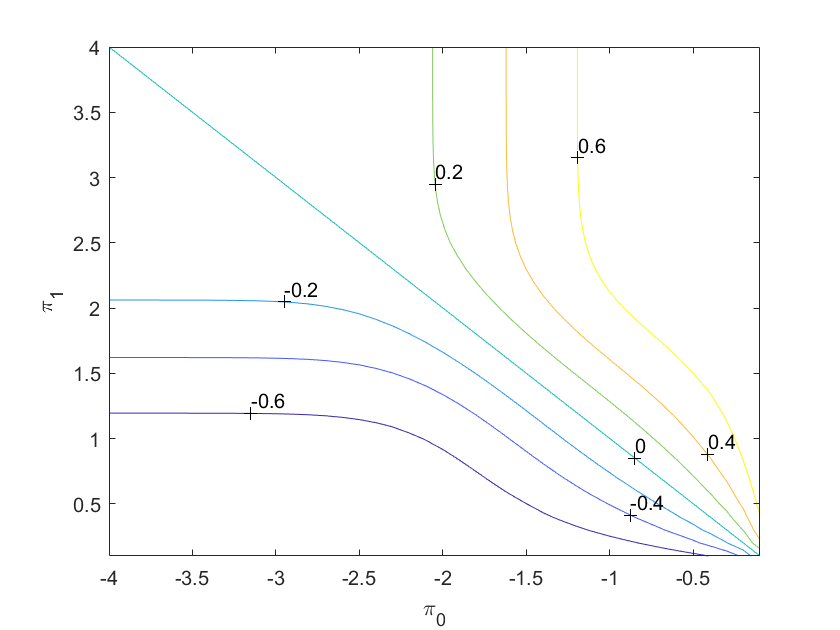}}
\end{center}
\par
\vspace{-10pt}
\caption{ The Sharpe Ratio as a function of $(\protect\underline{\protect\pi 
},\overline{\protect\pi })$ for $\protect\theta =0.0$ a)-b) $T=1.96$, c)-d) $%
T=4.26$, e)-f) $T=6.56$.}
\label{fig:theta=0.0}
\end{figure}

\begin{figure}[tbp]
\begin{center}
\subfloat[]{\includegraphics[width=0.5\textwidth]
{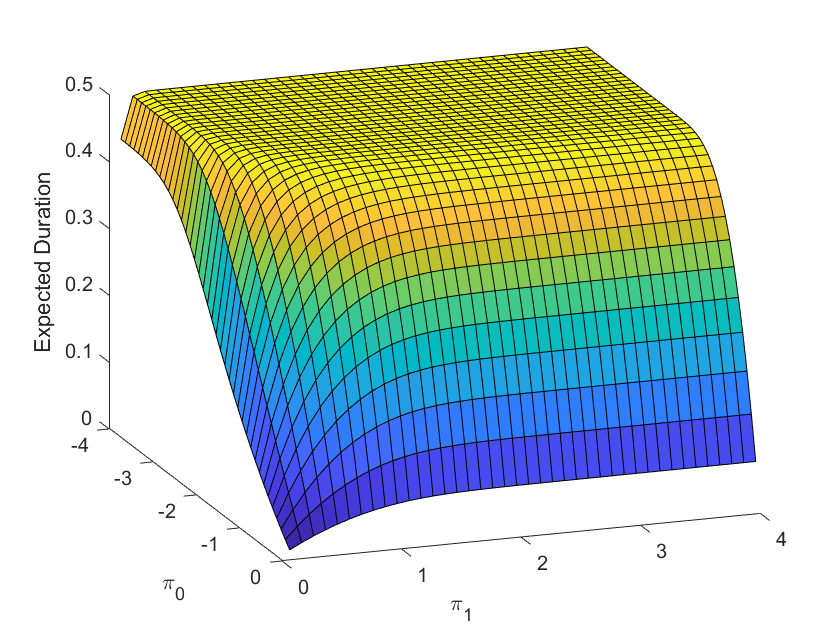}}
\subfloat[]{\includegraphics[width=0.5\textwidth]
{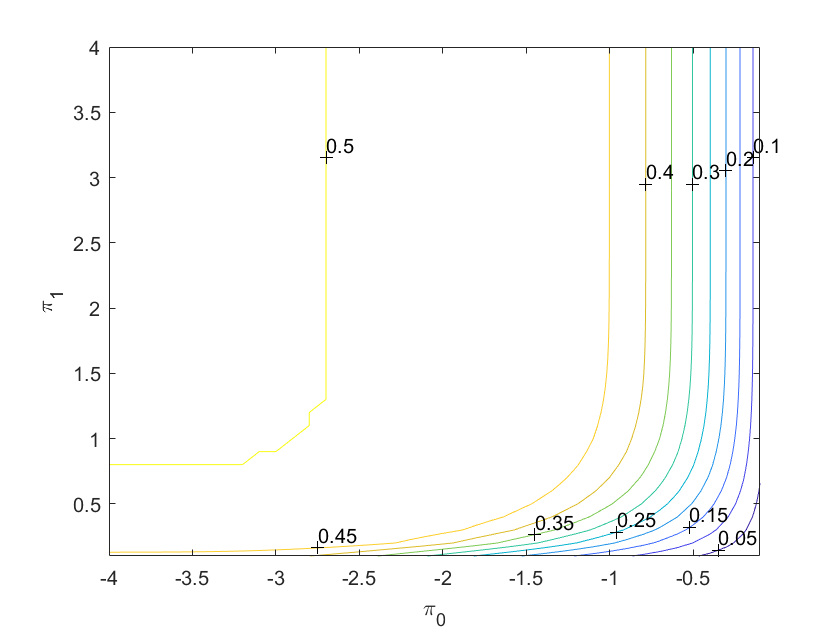}}%
\\[0pt]
\subfloat[]{\includegraphics[width=0.5\textwidth]
{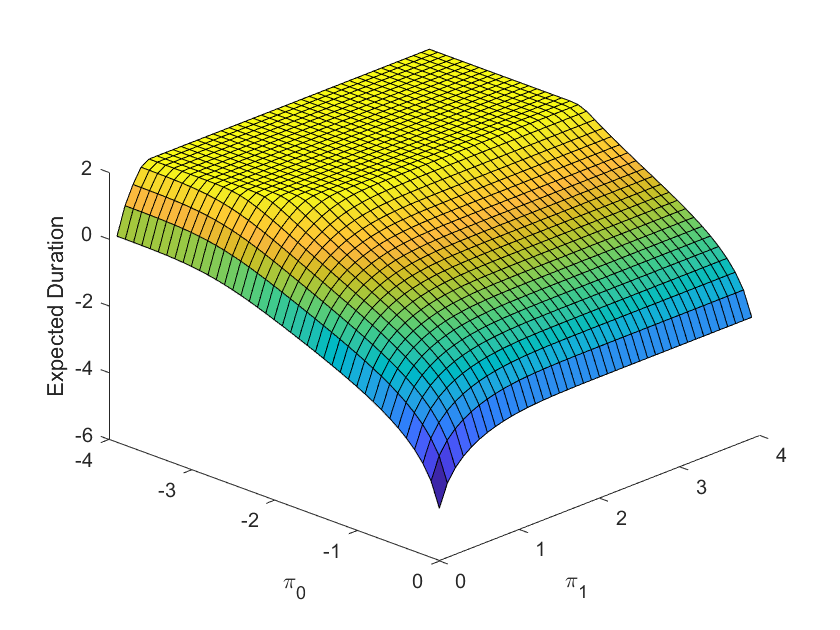}}
\subfloat[]{\includegraphics[width=0.5\textwidth]
{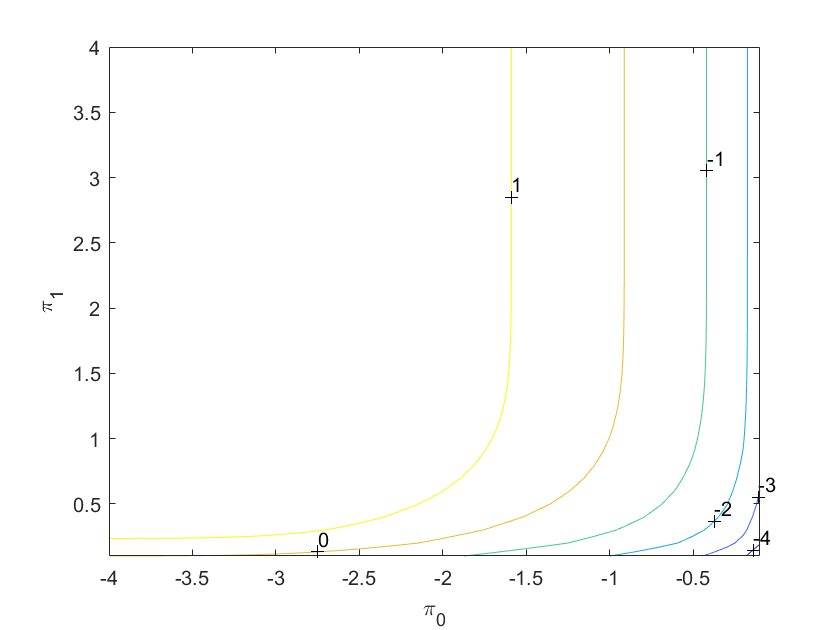}}
\end{center}
\par
\vspace{-10pt}
\caption{In Figures (a)-(b) we show the expected duration $\Upsilon =\left(
1-\exp \left( -2G\right) /2\right) $ as a function of $\protect\underline{%
\protect\pi }$, $\overline{\protect\pi }$; in Figures (c)-(d) we show the
logarithm of the expected duration $G$. The corresponding $\protect\theta %
=1.0$. Here and in Figures \protect\ref{fig:theta=1.0}, \protect\ref%
{fig:theta=0.5}, \protect\ref{fig:theta=0.0} $\protect\pi _{0}\equiv \protect%
\underline{\protect\pi }$, $\protect\pi _{1}\equiv \overline{\protect\pi }$.}
\label{fig:ed}
\end{figure}

\begin{figure}[tbp]
\begin{center}
\subfloat[]{\includegraphics[width=0.8\textwidth]
{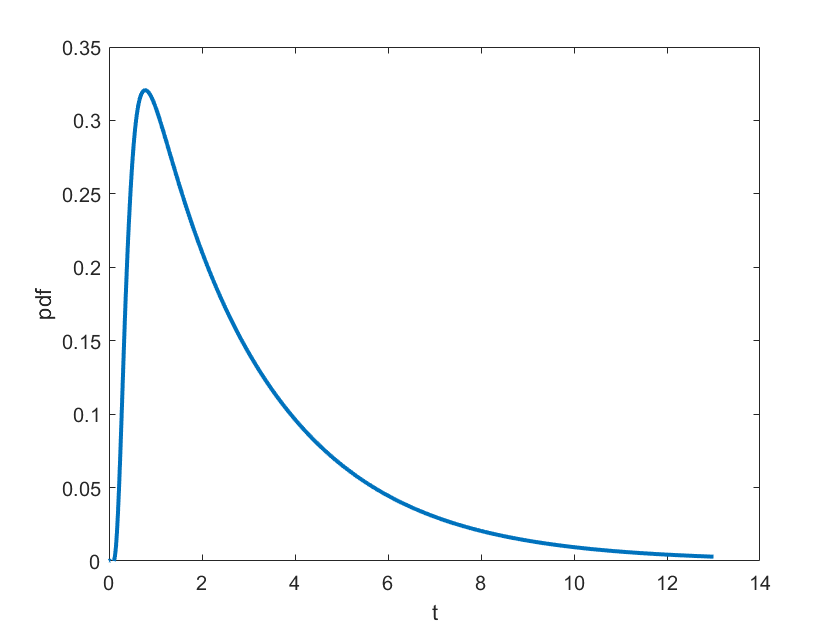}}
\\[0pt]
\subfloat[]{\includegraphics[width=0.8\textwidth]
{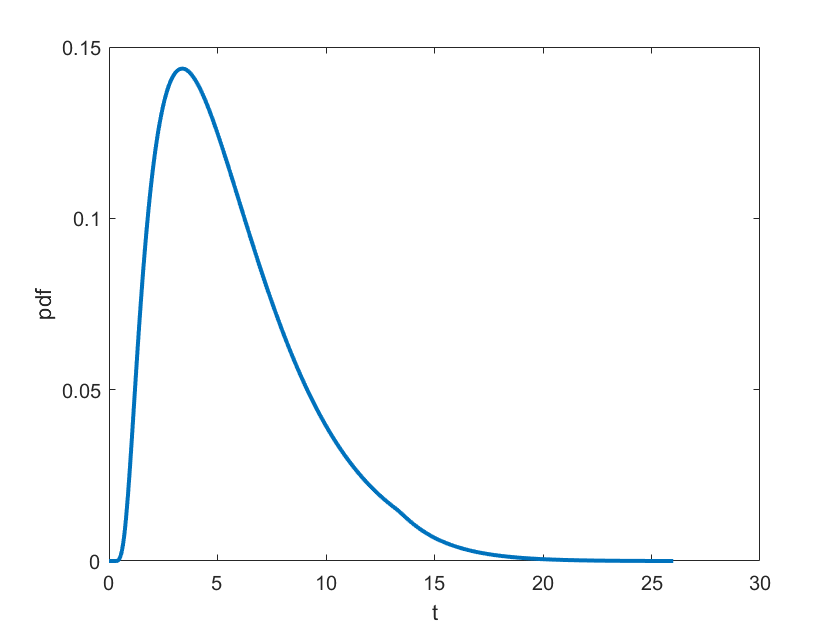}}
\end{center}
\par
\vspace{-10pt}
\caption{Figure (a) shows the pdf for the process $l\rightarrow u$; Figure
(b) shows the pdf for the round-trip process $l\rightarrow u\rightarrow l$.
The corresponding $l=-1/\protect\sqrt{2}$, $u=1/\protect\sqrt{2}$. We make
this choice because one standard deviation of the stationary O-U
distribution is $1/\protect\sqrt{2}$.}
\label{Fig12}
\end{figure}

\begin{figure}[tbp]
\begin{center}
\subfloat[]{\includegraphics[width=0.8\textwidth]
{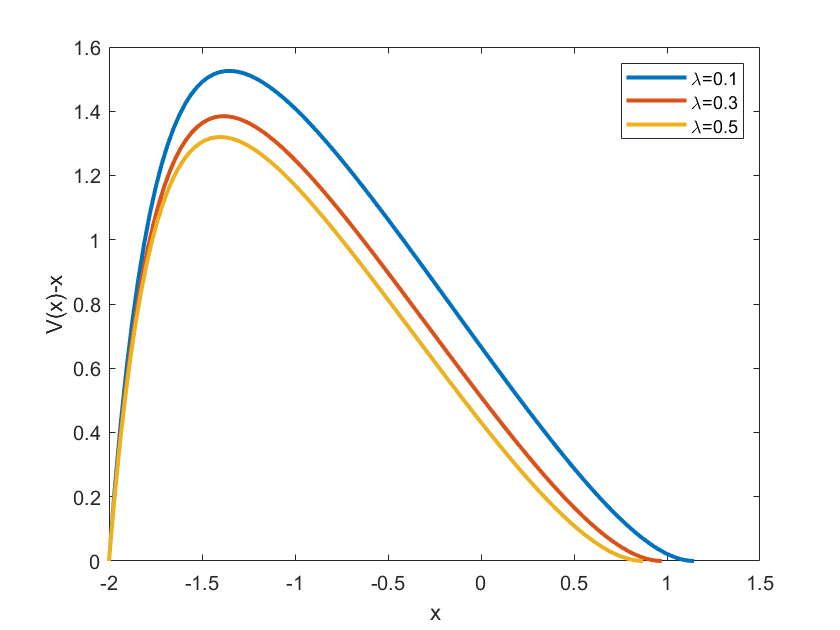}}
\\[0pt]
\subfloat[]{\includegraphics[width=0.8\textwidth]
{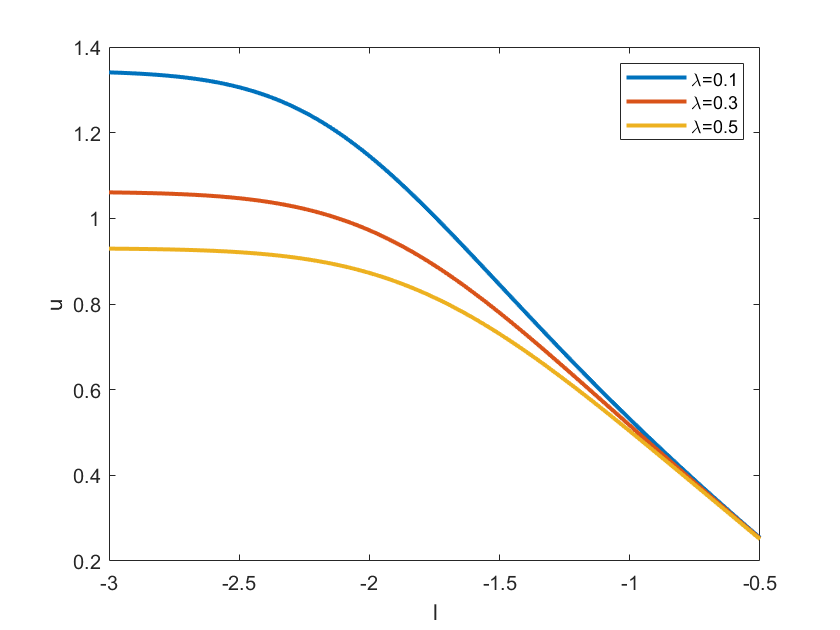}}
\end{center}
\par
\vspace{-10pt}
\caption{Figure (a) shows the nondimensional value function $V\left(
x\right) -x$ for several representative values of $\protect\lambda $, the
corresponding optimal values of $u$ are $1.15$, $0.97$, $0.87$; Figure (b)
shows the nondimensional optimal take-profit boundary $u$ as a function of
the nondimensional stop-loss boundary $l$.}
\label{Fig7}
\end{figure}

\begin{figure}[tbp]
\begin{center}
\subfloat[]{\includegraphics[width=0.8\textwidth]
{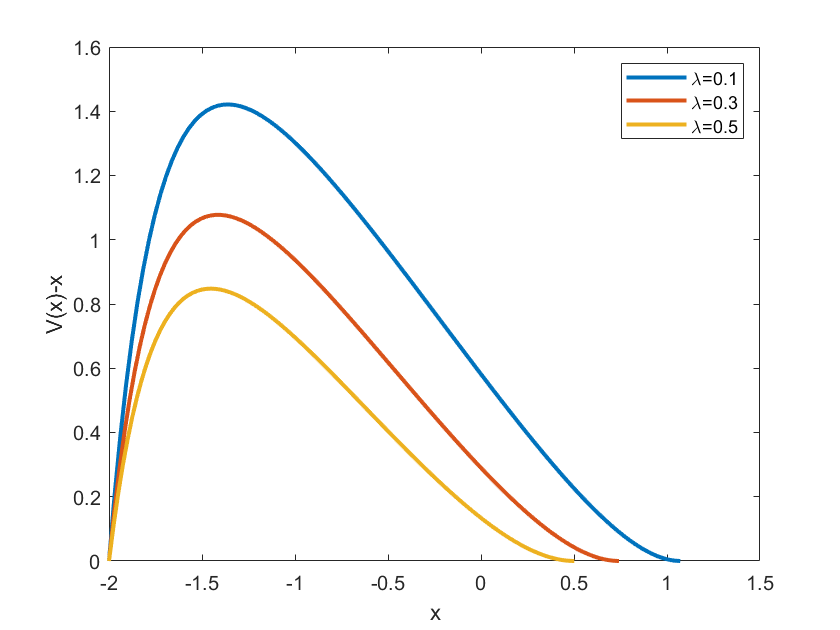}}
\\[0pt]
\subfloat[]{\includegraphics[width=0.8\textwidth]
{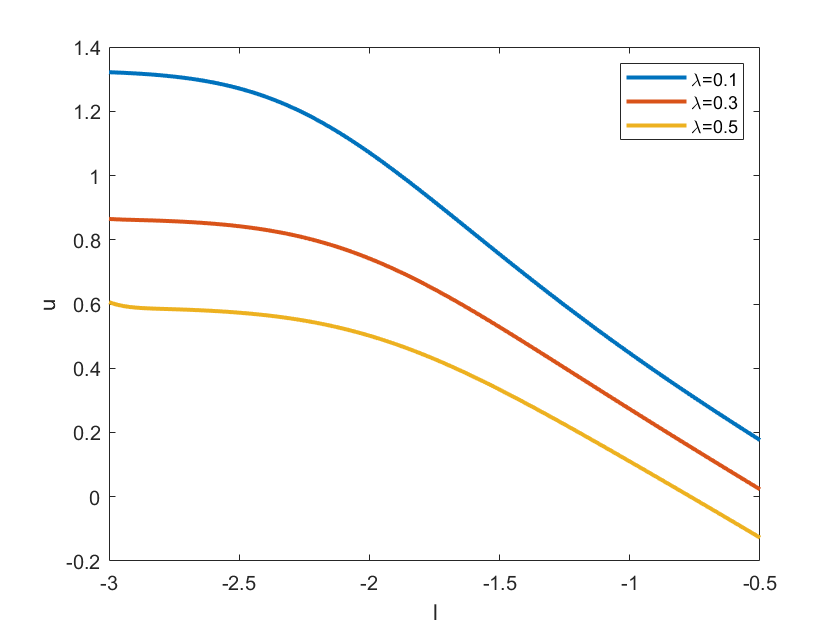}}
\end{center}
\par
\vspace{-10pt}
\caption{Figure (a) shows the nondimensional value function $V\left(
x\right) -x$ for $l=-2.0$, and several representative values of $\protect%
\lambda $, the corresponding optimal values of $u$ are $1.07$, $0.74$, $0.50$%
; Figure (b) shows the nondimensional optimal execution boundary $u\left(
l\right)$.}
\label{Fig8}
\end{figure}

\begin{figure}[tbp]
\begin{center}
\subfloat[]{\includegraphics[width=0.5\textwidth]
{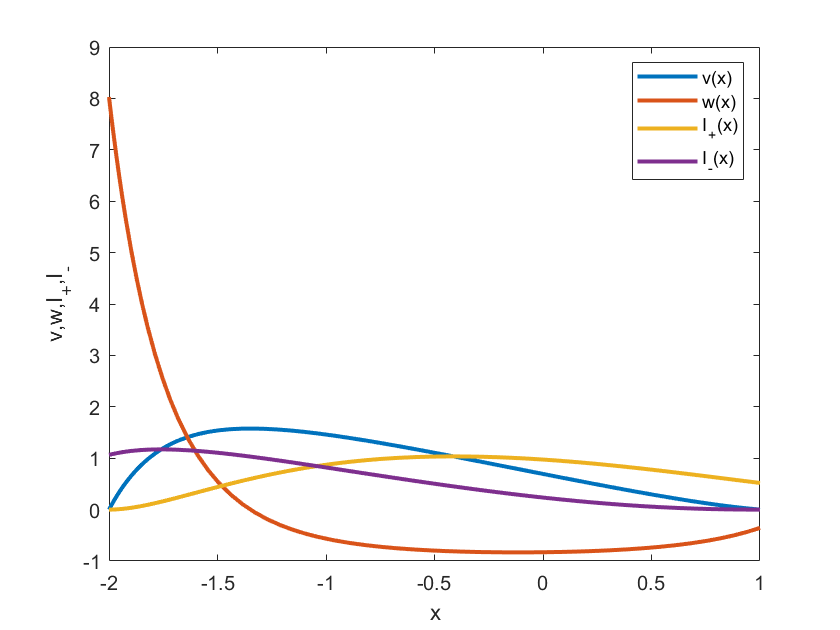}}
\\[0pt]
\subfloat[]{\includegraphics[width=0.5\textwidth]
{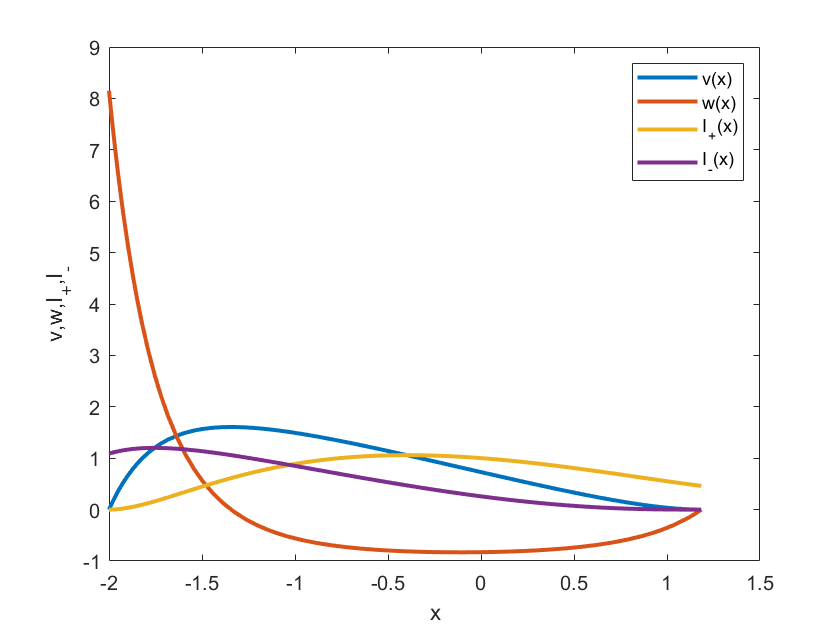}}
\\[0pt]
\subfloat[]{\includegraphics[width=0.5\textwidth]
{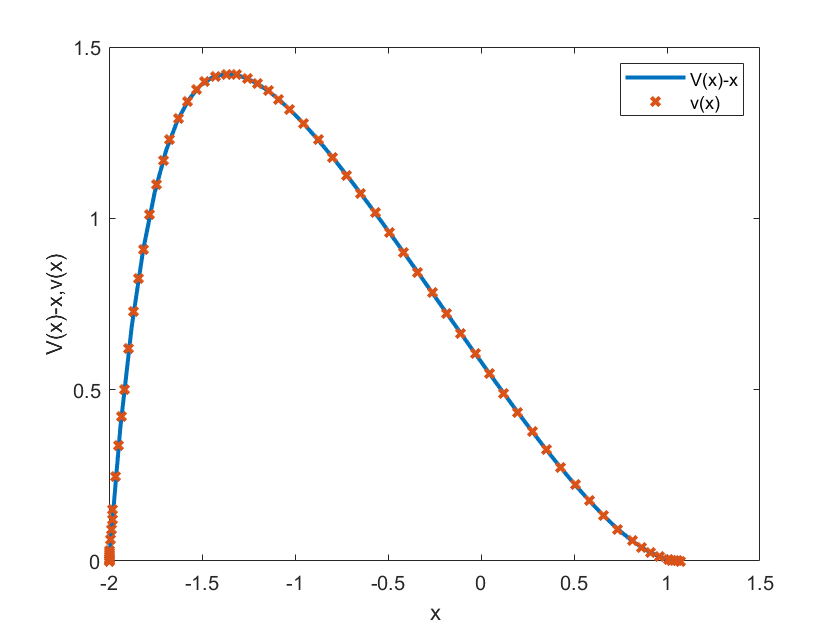}}
\end{center}
\par
\vspace{-10pt}
\caption{Figure (a) shows the nondimensional value functions $v\left(
x\right) $ for $l=-2.0$, $\protect\lambda =0.1$, $\protect\omega =0.1$, $%
\protect\kappa =1.0$. We choose $u=1$, which is not optimal. Hence, the
matching condition is not satisfied. Figure (b) shows the value function $%
v\left( x\right) $ for the optimal value of $u=1.18$. Since $u$ is optimal,
the matching condition is met, so that $w\left( u\right) =0$. Figure (c)
shows the nondimensional value functions $v\left( x\right) $ for $l=-2.0$, $%
u=1.18$, $\protect\lambda =0.1$, $\protect\omega =0.0$, $\protect\kappa =1.0$%
.}
\label{Fig9}
\end{figure}

\begin{figure}[tbp]
\begin{center}
\subfloat[]{\includegraphics[width=0.8\textwidth]
{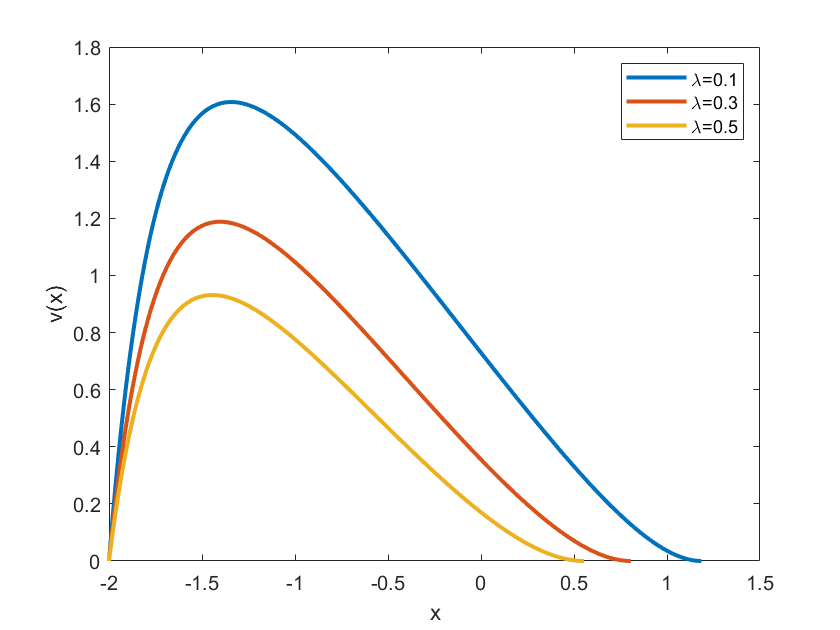}}
\\[0pt]
\subfloat[]{\includegraphics[width=0.8\textwidth]
{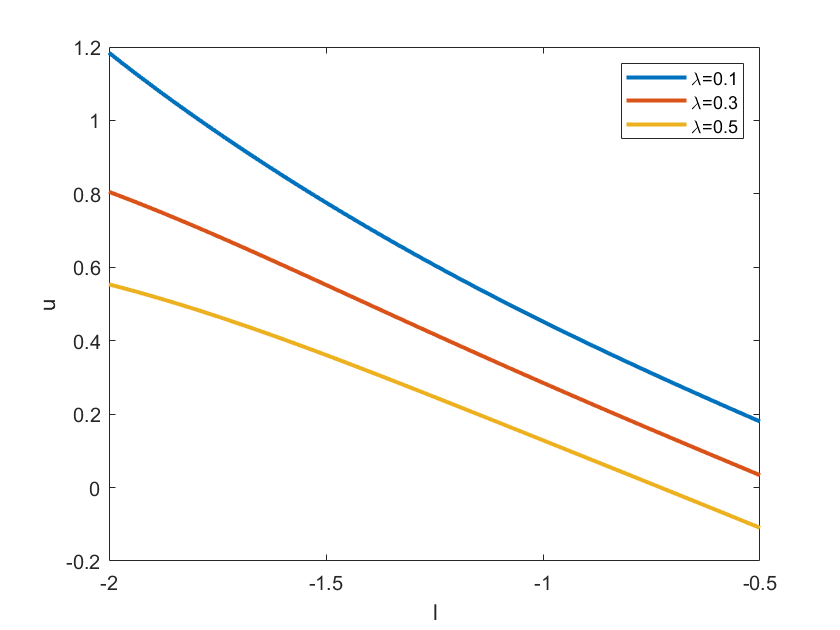}}
\end{center}
\par
\vspace{-10pt}
\caption{Figure (a) shows the nondimensional value functions $v\left(
x\right) $ for $l=-2.0$, and several representative values of $\protect%
\lambda$, the corresponding optimal values of $u$ are $1.18$, $0.81$Figure
(b) shows the nondimensional optimal execution boundary $u\left( l\right)$.
Here $\protect\omega =0.1$, $\protect\kappa =1.0$.}
\label{Fig10}
\end{figure}

\begin{figure}[h]
\begin{center}
\subfloat[]{\includegraphics[width=0.5\textwidth]
{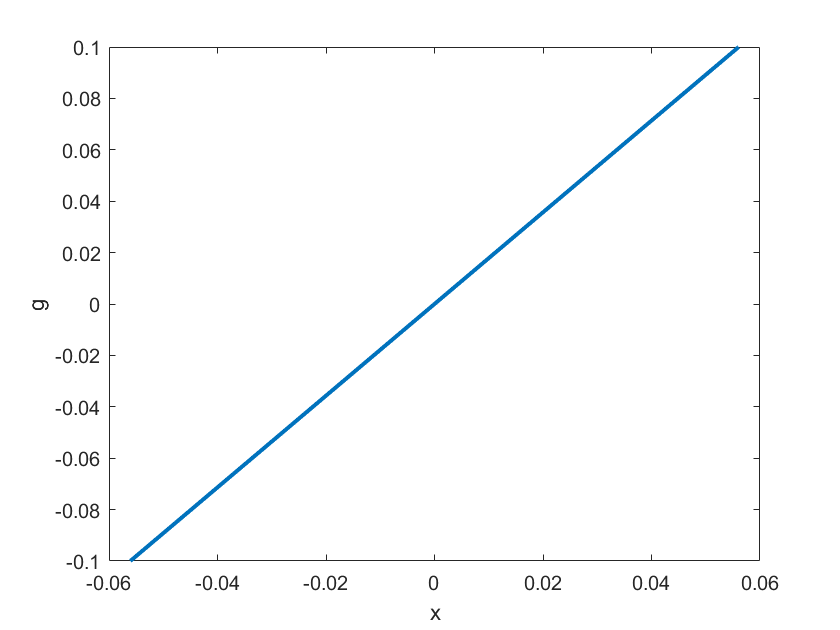}}
\\[0pt]
\subfloat[]{\includegraphics[width=0.5\textwidth]
{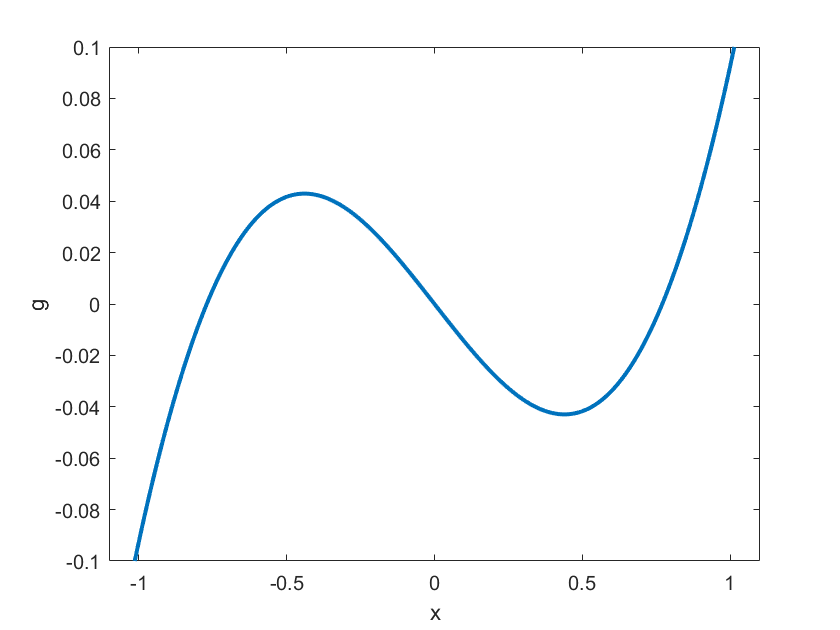}}
\\[0pt]
\subfloat[]{\includegraphics[width=0.5\textwidth]
{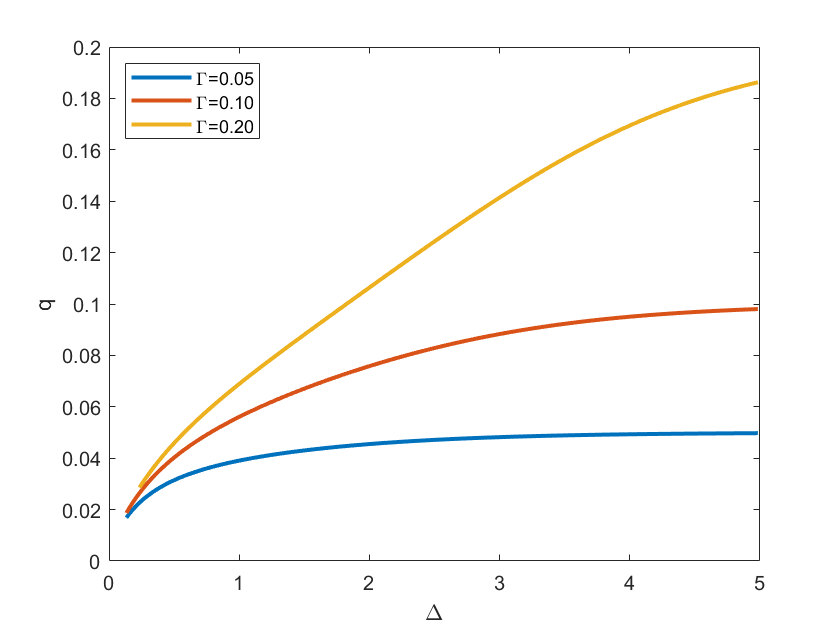}}
\end{center}
\par
\vspace{-10pt}
\caption{ In Figure (a) we show $g\left( x\right) $ corresponding to the
critical value $q=0.0561$; in Figure (b) we show $g\left( x\right) $
corresponding to the critical value $q=1.0131$. We can see that in the fist
case $g\left( x\right) $ has a single root at $x=0$, while in the second
case, there are three roots. In Figure (c) we show critical boundaries $%
q\left( \Delta \right) $ corresponding to three representative values of $%
\Gamma $, $\Gamma =0.05$, $0.10$, $0.20$. It is clear that for larger values
of $\Gamma $, it is beneficial to wait longer before changing one's position.
}
\label{Fig11}
\end{figure}

First, we present a method for calibrating the corresponding O-U process to
market prices. Second, we derive an explicit expression for the $\SR$ given
by Eq. (\ref{sharpe_final_eq}), and maximize it with respect to the stop
loss/ take profit bounds $(\underline{\pi },\overline{\pi })$. Third, for
three representative values of $\theta $, we calculate the $\SR$ on a grid
of $(\underline{\pi },\overline{\pi })$ and pre-chosen times and graphically
summarize in Figures \ref{fig:theta=1.0}, \ref{fig:theta=0.5}, \ref%
{fig:theta=0.0}. Next, for each case, we perform optimization and present $(%
\underline{\pi }^{\ast },\overline{\pi }^{\ast })$ in Table \ref{Tab1}. In
agreement with intuition, in the case of strong misprising, it is optimal to
wait until the trade's expiration without imposing stop losses/ take profit
bounds. For weaker mispricing, it is not optimal to stop losses, but it
might be optimal to take profits early. Still, to be on the safe side, we
recommend imposing stop losses chosen in accordance with one's risk appetite
to avoid unpleasant surprises caused by the misspecification of the
underlying process.

Our rules help liquidity providers to decide how to offer liquidity to the
market in the most profitable way, as well as by statistical arbitrage
traders to optimally execute their trading strategies.

A very interesting and difficult multi-dimensional version of these rules
(covering several correlated stocks) will be described elsewhere.

\begin{acknowledgement}
We greatly appreciate valuable discussions with Dr. Marsha Lipton, our
partner at Investimizer.
\end{acknowledgement}

\begin{acknowledgement}
We are grateful to Dr. Vadim Kaushansky for his help with an earlier version
of this paper.
\end{acknowledgement}

\newpage

\begin{table}[tbp]
\begin{center}
\begin{tabular}{|c|c|c|c|}
\hline
$\theta $ $\backslash $ $\Upsilon ,T$ & $0.49,\ 0.8$ & $0.4999,\ 1.96$ & $%
0.499999,\ 4.26$ \\ \hline
$1.0$ & 
\begin{tabular}{c}
$\underline{\pi }^{\ast }=-4.0$ \\ 
$\overline{\pi }^{\ast }=4.0$ \\ 
$\SR=1.2261$%
\end{tabular}
& 
\begin{tabular}{c}
$\underline{\pi }^{\ast }=-4.0$ \\ 
$\overline{\pi }^{\ast }=4.0$ \\ 
$\SR=1.3824$%
\end{tabular}
& 
\begin{tabular}{c}
$\underline{\pi }^{\ast }=-4.0$ \\ 
$\overline{\pi }^{\ast }=4.0$ \\ 
$\SR=1.3709$%
\end{tabular}
\\ \hline
$0.5$ & 
\begin{tabular}{c}
$\underline{\pi }^{\ast }=-4.0$ \\ 
$\overline{\pi }^{\ast }=0.6$ \\ 
$\SR=0.8219$%
\end{tabular}
& 
\begin{tabular}{c}
$\underline{\pi }^{\ast }=-4.0$ \\ 
$\overline{\pi }^{\ast }=0.9$ \\ 
$\SR=0.8792$%
\end{tabular}
& 
\begin{tabular}{c}
$\underline{\pi }^{\ast }=-4.0$ \\ 
$\overline{\pi }^{\ast }=1.0$ \\ 
$\SR=0.8963$%
\end{tabular}
\\ \hline
$0.0$ & 
\begin{tabular}{c}
$\underline{\pi }^{\ast }=-4.0$ \\ 
$\overline{\pi }^{\ast }=0.1$ \\ 
$\SR=0.7075$%
\end{tabular}
& 
\begin{tabular}{c}
$\underline{\pi }^{\ast }=-4.0$ \\ 
$\overline{\pi }^{\ast }=0.4$ \\ 
$\SR=0.7139$%
\end{tabular}
& 
\begin{tabular}{c}
$\underline{\pi }^{\ast }=-4.0$ \\ 
$\overline{\pi }^{\ast }=0.1$ \\ 
$\SR=0.7411$%
\end{tabular}
\\ \hline
\end{tabular}%
\end{center}
\caption{The Sharpe Ratio maximized over $(\protect\underline{\protect\pi },%
\overline{\protect\pi })$ for fixed $\Upsilon $ or $T$.}
\label{Tab1}
\end{table}

\end{document}